\documentclass[prd,onecolumn,nofootinbib,aps,tightenlines,preprintnumbers,notitlepage,longbibliography,superscriptaddress,12pt]{revtex4-1}

\usepackage[dvipsnames]{xcolor}
\usepackage{color}
\usepackage{tikz}
\usepackage[caption=false]{subfig}
\usepackage{hyperref}
\usepackage{multirow}
\usepackage{siunitx}
\usepackage[export]{adjustbox}% http://ctan.org/pkg/adjustbox
\usepackage{graphicx}
\usepackage{dcolumn}
\usepackage{bm}
\usepackage[normalem]{ulem}
\usepackage{epstopdf}
\usepackage{tabularx}
\usepackage{suffix}
\usepackage{mathtools}

\newcommand{\bn}{\hat{\boldsymbol{n}}}

\usepackage{graphicx}
\usepackage{dcolumn}
\usepackage{bm}
\usepackage[normalem]{ulem}
\usepackage{xcolor}

\graphicspath{ {figures/} }

\usepackage{epstopdf}
\newcommand{\be}{\begin{eqnarray}}
\newcommand{\non}{\nonumber \\}
\newcommand{\ee}{\end{eqnarray}}

\newcommand{\bbe}{\boldsymbol{e}}
\newcommand{\bv}{\boldsymbol{v}}

\newcommand{\bk}{\mathbf{k}}

\usepackage[export]{adjustbox}% http://ctan.org/pkg/adjustbox

\def\vl{\boldsymbol{\ell}}

\def\vL{\boldsymbol{L}}

\def\thetahat{\hat{\boldsymbol{\theta}}}
\def\phihat{\hat{\boldsymbol{\phi}}}
\def\bk{\boldsymbol{k}}

\def\spinup{\partial\kern-0.3em\raise0.42ex\hbox{\tiny\textbackslash}}
\def\spindown{\overline{\partial\kern-0.3em\raise0.42ex\hbox{\tiny\textbackslash}}}

\newcommand{\aap}{Astron. Astrophys.}

\newcommand{\mnras}{Mon. Not. R. Astron. Soc.}
\newcommand{\jcap}{J. Cosm. Astropart. Phys.}

\newcommand{\dd}{{\rm d}}

\newcommand{\jhu}{William H. Miller III Department of Physics and Astronomy, Johns Hopkins University, Baltimore, MD 21218, USA}

\newcommand{\illa}{Astronomy Department, University of Illinois at Urbana-Champaign, 1002 W. Green Street, Urbana, IL 61801, USA}

\newcommand{\illb}{Department of Physics, University of Illinois Urbana-Champaign, 1110 W. Green Street, Urbana, IL 61801, USA}

\newcommand{\perimeter}{Perimeter Institute for Theoretical Physics, 31 Caroline St N, Waterloo, ON N2L 2Y5, Canada}

\newcommand{\york}{Department of Physics and Astronomy, York University, Toronto, ON M3J 1P3, Canada}

\begin{document}

\title{Cosmology from the kinetic polarized Sunyaev Zel'dovich effect}

\author{Selim~C.~Hotinli}
\affiliation{\jhu}

\author{Gilbert~P.~Holder}
\affiliation{\illa}
\affiliation{\illb}

\author{Matthew~C.~Johnson}
\affiliation{\perimeter}
\affiliation{\york}

\author{Marc~Kamionkowski}
\affiliation{\jhu}

\date{\today}

%%%%%%%%%%%%%%%%%%%%%%
%%%% Abstract%%%%%%%%%
%%%%%%%%%%%%%%%%%%%%%%

\begin{abstract}
    
The cosmic microwave background (CMB) photons that scatter off free electrons in the large-scale structure induce a linear polarization pattern proportional to the remote CMB temperature quadrupole observed in the electrons' rest frame. The associated blackbody polarization anisotropies are known as the polarized Sunyaev Zel'dovich (pSZ) effect. Relativistic corrections to the remote quadrupole field give rise to a non-blackbody polarization anisotropy proportional to the square of the transverse peculiar velocity field; this is the kinetic polarized Sunyaev Zel’dovich (kpSZ) effect. In this paper, we forecast the ability of future CMB and galaxy surveys to detect the kpSZ effect, finding that a statistically significant detection is within the reach of planned experiments. We further introduce a quadratic estimator for the square of the peculiar velocity field based on a galaxy survey and CMB polarization. Finally, we outline how the kpSZ effect is a probe of cosmic birefringence and primordial non-Gaussianity, forecasting the reach of future experiments.
    
\end{abstract}

\maketitle

%%%%%%%%%%%%%%%%%%%%%%
%%%% Intro %%%%%%%%%%%
%%%%%%%%%%%%%%%%%%%%%%

\section{Introduction}

The next generation of CMB surveys such as the ongoing Simons Observatory~\citep{Ade:2018sbj,Abitbol:2019nhf}, the upcoming CMB-S4~\citep{Abazajian:2016yjj,Abazajian:2022nyh}, and  the futuristic CMB-HD~\citep{Sehgal:2019ewc,CMB-HD:2022bsz}, and galaxy surveys such as DESI~\citep{Aghamousa:2016zmz}, the Vera Rubin Observatory (VRO)~\citep{2009arXiv0912.0201L}, and perhaps then MegaMapper \citep{Schlegel:2019eqc} promise to greatly expand our ability to infer the fundamental characteristics of the Universe. One exciting new opportunity is to use the CMB as a back-light, observing the effects on the CMB of scattering and lensing by the intervening large-scale structure. These interactions include the thermal, kinetic and polarized Sunyaev Zel’dovich effects~\citep{1969Ap&SS...4..301Z,1970A&A.....5...84Z,Sunyaev:1980vz,1972CoASP...4..173S,1999MNRAS.310..765S,itoh1998relativistic,Challinor:1999yz,Kamionkowski:1997na,Chluba:2012py}, and the
integrated Sachs-Wolfe effects~\citep{1967ApJ...147...73S}, which includes the
moving-lens effect~\citep{1983Natur.302..315B,2002PhRvD..65h3518C,Hotinli:2018yyc,Hotinli:2020ntd}, and weak-gravitational lensing (for a review see e.g.~\citep{Lewis:2006fu}). Each of these effects can provide new insights and constraints on cosmological models. 

When CMB photons scatter from electrons, they acquire a specific pattern of linear polarization proportional to the remote CMB temperature quadrupole seen by the electron in its rest frame. The dominant contributions to the remote quadrupole field are primordial in nature, arising from the Sachs Wolfe and Doppler effects at decoupling. In the late-Universe, the next most significant contribution is from the integrated Sachs Wolfe effect. These contributions all give rise to a blackbody polarization signal, and are described in detail in e.g.~\citep{Deutsch:2017cja}. Primordial gravitational waves can also contribute to the remote quadrupole field~\citep{Alizadeh:2012vy,Deutsch:2017ybc,Deutsch:2018umo}, giving rise to an additional blackbody polarization signal. 

Another contribution to the remote quadrupole field arises from the relativistic Doppler effect. This `kinematic' quadrupole ~\citep{Sunyaev:1980vz,1972CoASP...4..173S,Baumann:2002es,Cooray:2002cb,2000ApJ...529...12H,Shimon:2006hn,Bunn:2006mp,Yasini:2016pby} gives rise to a non-blackbody polarization anisotropy, which we refer to as the kinematic polarized Sunyaev Zel'dovich (kpSZ) effect, whose study is the focus of this work~\footnote{As described in detail in Ref.~\cite{Yasini:2016pby}, there are additional frequency-dependent secondary polarization anisotropies associated with the relativistic mixing of the primordial dipole and octupole into the locally observed quadrupole of a moving scatterer. These effects are generally smaller than the pSZ and kpSZ effects considered here, although they can dominate in specific regions of the sky and at high frequencies~\cite{Yasini:2016pby}.}. While subdominant to the polarization anisotropies associated with recombination, reionization, and the pSZ effect, the kpSZ effect has a unique frequency dependence which allows it to be separated using multi-frequency CMB experiments. The kpSZ effect has been studied previously as a signal in e.g.~Refs.~\citep{Roebber:2013lra,Yasini:2016pby,Fidler:2017irr} and as a bias on the pSZ measurement in~e.g.~Ref.~\citep{Hall:2014wna}. Here, we provide a detailed calculation of the $E$- and $B$-mode signals induced by the kpSZ up to non-linear order and assess the prospects to detect this signal through cross-correlating the CMB with templates of the kinematic quadrupole that can be reconstructed from measurements of galaxies. We highlight two distinct methods for detection: first, through cross-correlating the CMB maps on degree scales with kpSZ templates from galaxy surveys, and next, through reconstructing the remote kinetic quadrupole with pSZ tomography~\citep{2012PhRvD..85l3540A,Deutsch:2018umo,Deutsch:2017ybc,Deutsch:2017cja,lavaux2004sunyaev,amblard2005sunyaev}. The latter technique has been shown to be a promising direction for multitudes of cosmological analyses, adding to the prospects to detect gravitational tensor perturbations~\cite{Deutsch:2018umo} and improving the constraints on the mean optical depth to reionization~\cite{Meyers:2017rtf}, for example. 

More generally, the reconstruction of large-scale cosmological fluctuations via cross-correlating imprints of scattering and lensing effects on the CMB with tracers of structure~\cite{Dvorkin:2008tf,Dvorkin:2009ah,Smith:2016lnt,Kamionkowski:2008fp,Yadav:2009eb,Gluscevic:2009mm,Alizadeh:2012vy,Deutsch:2017cja,Deutsch:2017ybc,Meyers:2017rtf,Hotinli:2018yyc,Hotinli:2020ntd,Hotinli:2021hih,Hotinli:2020csk,Cayuso:2021ljq} has a wide range of applications. These include the reconstruction of large-scale velocity fields of matter via measurements of the kinetic Sunyaev-Zel'dovich (kSZ) effect~\citep{1969Ap&SS...4..301Z,1970A&A.....5...84Z,1980ARA&A..18..537S,1972CoASP...4..173S,Sazonov:1999zp}, which is anticipated to provide competitive constraints on local-type primordial non-Gaussianity~\cite{Deutsch:2017ybc,Smith:2018bpn,Munchmeyer:2018eey}, deviations from general relativity~\cite{Zhang:2015uta}, specific forms of primordial isocurvature~\cite{Hotinli:2019wdp}, sources of CMB anomalies~\cite{Cayuso:2019hen}, and the reionization history~\cite{Alvarez:2020gvl,Hotinli:2020csk}. Here, we focus on the kpSZ effect {and describe possible ways to utilize this signal for cosmological inference.} 

We begin in Sec.~\ref{sec:signal} by reproducing the calculation of the kpSZ effect, presenting the $E$- and $B$-mode polarization signal, including contributions from the inhomogeneous optical depth at low redshift. In Sec.~\ref{sec:detection} we forecast the ability of future CMB experiments to detect the kpSZ effect using two techniques. First, we demonstrate that on large angular scales the kpSZ effect can be detected by cross-correlating CMB polarization with a template constructed using a Rubin LSST-like galaxy survey. Next, we introduce a quadratic estimator for the square of the transverse velocity field based on the statistically anisotropic cross-correlation of CMB polarization with a galaxy survey. Either technique can be used to obtain a detection with signal-to-noise greater than 2 using CMB-S4, with improvements expected for experiments with greater resolution (e.g. CMB-HD) or additional frequency channels (e.g. PIXIE). In Sec.~\ref{sec:cosmology}, we demonstrate that the kpSZ effect can be used to constrain cosmic birefringence and primordial non-Gaussianity, with constraints that scale well with noise and resolution of future experiments. We conclude in Sec.\ref{sec:conclusions}. 

\section{Kinetic Polarized Sunyaev Zel'dovich Signal}\label{sec:signal}

\subsection{The kinetic quadrupole} 

Here, we calculate the kinetic quadrupole signal following similar calculations for the pSZ signal as in e.g.~Refs.~\citep{Meyers:2017rtf, Hall:2014wna}. The Stokes parameters $Q$ and $U$ that are measured in CMB observations in a coordinate frame centred on the observer can be projected onto a 2-sphere with unit vectors $\{\thetahat,-\phihat\}$ in the plane perpendicular to line of sight direction $\bn$, using polarization vectors $\bbe_\pm(\bn)=(\thetahat\mp i\phihat)/\sqrt{2}$. The resulting complex CMB polarization along the line of sight takes the form of an integral in comoving distance $\chi$,  
\be
(Q\pm iU)(\bn)=\int_{0}^{\chi_\star}\dd\chi \ \dot{\tau}(\chi\bn)e^{-\tau(\chi)}\,_{\pm}p(\chi\bn)\, , 
\ee
where $\chi$ is the comoving distance along our past light cone, $\chi = 0$ today, $\chi_*$ is the comoving distance to the surface of last scattering, and $\tau$ is the optical depth along the line of sight. We write the optical depth at redshift $z$ along $\bn$ as $\dot{\tau}(\chi\bn)=\dd\tau/\dd\chi(\chi\bn)$ where $\dd\tau/\dd\chi(\chi\bn)=\sigma_Tan_e(\chi\bn)$ and $\sigma_T$ is the Thomson
cross section, $a$ is the scale factor, $n_e$ is the free electron number density. The remote quadrupole field projected along the line of sight $\,_{\pm}p(\chi\bn)$ is given by:
\be
\,_{\pm}p(\chi\bn) \equiv \sum_{m=-2}^{m=2} \Theta_{2m}(\chi\bn) \,_{\mp2}Y_{\ell m}(\bn)   
\ee
where $\Theta_{2m}(\chi\bn)$ are the $\ell =2$ moments of the CMB temperature anisotropies observed at the position $\chi\bn$ along our past light cone. Although our perspective will mainly be that the optical depth and remote quadrupole are continuous fields, it is sometimes helpful to consider the contribution from a single scatterer (e.g. a cluster) with optical depth $\tau$
\be
(Q\pm iU)^{(e)}(\bn)= \tau \,_{\pm}p(\bn)\,,
\ee
where we omit the $\chi$ dependence in what follows. Isolating the quadrupole field
\be\label{eq:quad_source0}
\,_{\pm}p(\bn)=\frac{1}{\tau} (Q\pm iU)^{(e)}(\bn)\ , 
\ee
we can expand this into spin-2 spherical harmonics on a shell of constant $\chi$ as 
\be\label{eq:source_multipoles}
\,_{\pm}p(\bn)=\sum\limits_{\ell m}\,_{\mp}a^p_{\ell m}\,_{\mp2}Y_{\ell m}(\bn)\ . 
\ee
Note that $(Q + iU)(\bn)$ in our formalism is spin\ $-2$ due to the choice of coordinate frame centred at the observer with scattered photon travelling in the $-\bn$ direction.

For a scatterer at rest, the remote quadrupole field receives contributions from the Sachs Wolfe effect, primordial Doppler effect, integrated Sachs Wolfe effect, and potentially primordial gravitational waves. When the scatterer has a non-zero peculiar velocity, relativistic abberation in the rest frame of the scatterer will lead to frequency-dependent Doppler leakage of adjacent multipoles into the quadrupole. The dominant effect is the leakage of the monopole into the quadrupole (see Ref.~\cite{Yasini:2016pby} for  contributions from the dipole and octupole), which can be understood as follows.  In the electron rest frame, the CMB black-body temperature gets a relativistic correction as $T'=T f(x) (1-{\bv}_e\cdot\bn_e)/\sqrt{1-|\bv_e|^2}$ where ${\bv}_e$ is the electron peculiar velocity, $f(x)$ captures the frequency dependence of the effect, $T$ is the CMB black-body temperature at the CMB rest frame, where the aberration of the CMB fluctuations vanishes and we defined $\bn_e$ in observer's frame as the direction of the photon incoming on the electron.  

The relativistic correction to the quadrupole can be calculated as equal to $-(1/10)\,g(x)\beta_\perp^2$ where $\beta_\perp^2=|\bv_e|^2(1-\bn_e\cdot\bn)$, $\beta_\perp$ is the transverse velocity amplitude \footnote{Note that the longitudinal velocity contribution will get a relativistic correction $v/c$ when transformed back to CMB frame and is sub-dominant.}\!, $g(x)=(x/2)\coth(x/2)$ and $x=h\nu/k_BT$ is the dimensionless frequency of the photon in CMB rest frame. 

We can write the kinetic contribution to the remote quadrupole field as 
\be\label{eq:quad_source_kin}
{\,_{\pm}}p(\chi\bn) = -\frac{T_{\rm CMB}}{10}\left(\frac{x}{2}\coth{\frac{x}{2}}\right)\left(V_\theta (\chi\bn) \mp iV_\phi (\chi\bn)\right)^2\,,
\ee
where we projected the velocity transverse to the line of sight ${\bv}_\perp(\bn)$ at each $\chi$, an ordinary 2-vector field in 2-sphere, onto the polarization basis we use $\bbe_\pm(\bn)$, with definitions $V_\theta:={\bv}_\perp\cdot\thetahat$ and $V_\phi:={\bv}_\perp\cdot\phihat$. The transverse velocity can be decomposed into a gradient and a curl part as
\be\label{eq:grad_curl}
{\bv}_\perp(\bn)=\boldsymbol{\nabla}\Upsilon(\bn)+\boldsymbol{\nabla}\times\varpi(\bn). \ee
We define the commonly used spin raising and lowering operators as
\be\label{eq:spin_updown}
\spinup\,_s\eta & =-\sin^s\!\theta(\partial_\theta+i\mathrm{cosec}\theta\!\ \partial_\phi)(\sin^{-s}\!\theta\,_s\eta)\non 
\spindown\,_s\eta& =-\sin^s\!\theta(\partial_\theta{
-}i\mathrm{cosec}\theta\!\ \partial_\phi)(\sin^{-s}\!\theta\,_s\eta)\,,
\ee
which are useful to define higher spin harmonics in terms of standard spin-weighted spherical harmonics, for example,
\be
\,_1Y_{\ell m}&&=\ \ \ \spinup\left(\sqrt{\frac{(\ell-1)!}{(\ell+1)!}}Y_{\ell m}\right)\,, \\
\,_{-1}Y_{\ell m}&&=-\spindown\left(\sqrt{\frac{(\ell-1)!}{(\ell+1)!}}Y_{\ell m}\right)\, . 
\ee 
Moreover, using Eq.~\eqref{eq:grad_curl} and Eq.~\eqref{eq:spin_updown}, we can show that
\be
{\bv}_\perp\cdot\bbe_{-} && =-\spinup(\Upsilon-i\varpi)=v_\theta+iv_\phi\,, \\ {\bv}_\perp\cdot\bbe_{+} && =-\spindown(\Upsilon+i\varpi)=v_\theta-iv_\phi\, .
\ee
We can expand the fields $\Upsilon$ and $\varpi$ in spherical harmonics as
\be\label{eq:spin0_defn}
\Upsilon(\bn) &&= \sum\limits_{\ell m}\sqrt{\frac{(\ell-1)!}{(\ell+1)!}}V^G_{\ell m}Y_{\ell m}(\bn)\,, \\
\varpi(\bn) &&= \sum\limits_{\ell m}\sqrt{\frac{(\ell-1)!}{(\ell+1)!}}V^C_{\ell m}Y_{\ell m}(\bn)\, ,
\ee
which then lets us write 
\be\label{eq:spin1_exp}
(V_\theta\pm iV_\phi)(\bn)=\sum\limits_{\ell m}({{\pm}} V^G_{\ell m}+i V^C_{\ell m})(\bn)\,_{\pm1}Y_{\ell m}(\bn)\,.
\ee
Under parity transformation, the transverse velocity transforms as $\boldsymbol{v}(\bn)\rightarrow-\boldsymbol{v}(-\bn)$ while the transverse components transform as $(V_\theta\pm iV_\phi)(\bn)\rightarrow-(V_\theta\mp iV_\phi)(-\bn)$. Taking the parity transformation properties of the spin-1 harmonics, one finds that in order to satisfy the parity transformation properties of the velocity, the expansion coefficients must transform as $V^G_{\ell m} \rightarrow(-1)^{\ell} V^G_{\ell m}$ and $V^C_{\ell m}\rightarrow(-1)^{\ell+1} V^C_{\ell m}$, i.e. with electric and magnetic parities respectively. Finally, note that in the presence of only scalar perturbations, we expect only the gradient term to be present, i.e. we will approximate $V^C_{\ell m}\approx0$.

The transverse velocity potential $\Upsilon$ projected into the 2-sphere perpendicular to line-of-sight $\bn$ can be calculated from the velocity field ${\bv}(\chi\bn)$ by taking a projected derivative $\nabla_{\bn}$ of the velocity potential, and is related to the covariant derivative on the sphere as $\nabla_{\bn}=(1/\chi)\nabla$. The Fourier transform of $\Upsilon$ can then be written as
\be\label{eq:velo_pot}
\Upsilon(\chi\bn)=-\int\frac{\dd^3\bk}{(2\pi)^3}\frac{v(\bk,\chi)}{k\chi}e^{ik\chi\hat{\bk}\cdot\bn}\,,
\ee 
where $v({\bk,\chi})=\hat{\boldsymbol{k}}\cdot{\bv}(\chi\bn)$, and we have reintroduced the conformal time dependence. The coefficient $V^G_{\ell m}$ defined above can be found by expanding the exponential in Eq.~\eqref{eq:velo_pot} as
\be
\exp(i k\chi\hat{\bk}\cdot{\bn})=4\pi\sum\limits_{\ell m}i^{\ell}j_\ell(k\chi)Y_{\ell m}(\bn)Y_{\ell m}^*(\hat{\bk})\, ,
\ee
and isolating the spherical harmonic $Y_{\ell m}({\bn})$, we get~\citep{Hall:2014wna,Bunn:2006mp}
\be\label{eq:pot_quad}
V^G_{\ell m}=-4\pi i^\ell\sqrt{\ell(\ell+1)}\int\frac{\dd^3 \bk}{(2\pi)^3}{v(\bk,{\chi})}\frac{j_\ell(k\chi)}{k\chi}Y_{\ell m}^*(\hat{\bk})\ .
\ee

We now use the expression for $V^G_{\ell m}$ to calculate the harmonic coefficient of the polarization sourced by the kinematic quadrupole and its spectrum $p_{\ell m}$. Using Eqs.~\eqref{eq:source_multipoles},~\eqref{eq:quad_source_kin}~and~\eqref{eq:spin1_exp}, we can write 
\be\label{eq:eq19}
\sum_{\ell m}\,_{\pm}p_{\ell m}\,_{\pm2}  Y_{\ell m}(\bn) = f(x)\sum_{\ell m}\sum_{\ell' m'}V^G_{\ell m}V^G_{\ell' m'}\,_{\pm1} Y_{\ell m}(\bn)\,_{\pm1} Y_{\ell' m'}(\bn)\, .
\ee
where $f(x)=-(1/10)g(x)T_{\rm CMB}$ { and we use notations as $\,_{\pm}p_{\ell m}=\,_{\pm}a_{\ell m}^p$ for brevity}. Some useful spherical harmonics relations are 
\be\label{eq:orth_1}
\int\dd^2\bn\,_sY_{\ell m}(\bn)\,_sY_{\ell' m'}^{\star}(\bn)=\delta_{\ell\ell'}\delta_{mm'}\,,
\ee
and 
\be\label{eq:orth_2}
\int\!\!\dd^2\bn\,_{s_1}\!Y_{\ell_1 m_1}(\bn)\,_{s_2}\!Y_{\ell_2 m_2}&&(\bn)\,_{s_3}Y_{\ell_3 m_3}(\bn)\\=
&&\sqrt{\frac{(2\ell_1+1)(2\ell_2+1)(2\ell_3+1)}{4\pi}}
\begin{pmatrix}
\ell_1 & \ell_2 & \ell_3 \\
  m_1 & m_2 & m_3 
\end{pmatrix}
\begin{pmatrix}
\ell_1 & \ell_2 & \ell_3 \\
 \!-s_1 & \!-s_2 & \!-s_3 
\end{pmatrix}\,. \nonumber
\ee
We multiply both sides { of Eq.~\eqref{eq:eq19}} by the spin-2 spherical harmonic $\,_{\mp2}Y_{L M}^*(\bn)$ and integrate over $\dd^2\bn$. Using the relations defined in Eqs.~\eqref{eq:orth_1}~and~\eqref{eq:orth_2} and $\,_{\mp2}Y_{LM}^*(\bn)=\,_{\mp2}Y_{L-M}(\bn)$, we get 
\be\label{eq:remote_quad}
\,_{\pm}p_{\ell m}\!=\! f(x)\sum_{\ell' m'}\sum_{\ell'' m''}V^G_{\ell' m'}V^G_{\ell'' m''}\,_{\pm}\mathcal{I}_{\ell'\ell''\ell}^{m'm''m}\,,
\ee
where we defined 
\be
\,_{\pm}\mathcal{I}_{\ell'\ell''\ell}^{m'm''m}=\sqrt{\frac{(2\ell'+1)(2\ell''+1)(2\ell+1)}{4\pi}}\begin{pmatrix}
  \ell' & \ell'' & \ell \\
  m' & m'' & -m' 
  \end{pmatrix}
  \begin{pmatrix}
  \ell' & \ell'' & \ell \\
  \pm1 & \pm1 & \mp2 
  \end{pmatrix}\,.
\ee
This equation suggests that the quadrupole coefficient $\,_{\pm}p_{LM}$ satisfies the reality condition, i.e. $X^*_{\ell m}=X_{\ell -m}$, if $V^{G\,*}_{\ell m}=V^{G}_{\ell -m}$ does as well.\footnote{Seeing this also requires the symmetry relation
\be
  \begin{pmatrix}
   \ell_1 & \ell_2 & \ell_3 \\
   m_1 & m_2 & m_3 
  \end{pmatrix}=(-1)^{\ell_1+\ell_2+\ell_3}
    \begin{pmatrix}
   \ell_1 & \ell_2 & \ell_3 \\
   -m_1 & -m_2 & -m_3 
  \end{pmatrix}
\ee
which returns a factor $(-1)^{2(\ell_1+\ell_2+\ell_3)}$ for $p^{*}_{\ell m}$.} 
The latter equality can be seen to be true by observing Eq.~\eqref{eq:spin0_defn} and noting that we take $V^C_{\ell m}=0$. We define the variance of the coefficients in the usual way, 
\be\label{eq:Palm}
\langle X_{\ell m}Y^{\star}_{\ell' m'}\rangle=C^{XY}_\ell\delta_{\ell\ell'}\delta_{mm'}\, . 
\ee

Next, we calculate the kinematic quadrupole auto-spectra $\langle p_{\ell m}p^{*}_{\ell' m'}\rangle$ which can be found by using the reality of $p_{LM}^{*}$ in order to satisfy Eq.~\eqref{eq:Palm} as
\be\label{eq:auto_corr}
\begin{split}
\langle\,_{\pm}p_{\ell m}\,_{\pm}p^{*}_{\ell'm'}\rangle&=f^2(x)\sum\limits_{\ell_1m_1}\sum\limits_{\ell_1'm_1'}\sum\limits_{\ell_2m_2}\sum\limits_{\ell_2'm_2'}\langle V^G_{\ell_1m_1}V^G_{\ell_1'm_1'}V^{G\,*}_{\ell_2m_2}V^{G\,*}_{\ell_2'm_2'}\rangle
\,\,_{\pm}\mathcal{I}_{\ell_1\ell_1'\ell}^{m_1m_1'm}
\,_{\pm}\mathcal{I}_{\ell_2\ell_2'\ell'}^{m_2m_2'm'}
\end{split}
\ee
where we collected the frequency dependent terms assuming equal frequencies. We write the four-point function defined in the second line in Eq.~\eqref{eq:auto_corr} as 
\be\label{eq:four_point}
\langle V^G_{\ell_1m_1}&&V^G_{\ell_1'm_1'}V^{G\,*}_{\ell_2m_2}V^{G\,*}_{\ell_2'm_2'}\rangle \\ 
&&=\langle V^G_{\ell_1m_1}V^G_{\ell_1'm_1'}\rangle \langle V^{G\,*}_{\ell_2m_2}V^{G\,*}_{\ell_2'm_2'}\rangle
+ \langle V^G_{\ell_1m_1}V^{G\,*}_{\ell_2'm_2'}\rangle \langle V^{G\,*}_{\ell_2m_2}V^G_{\ell_1'm_1'}\rangle
+ \langle V^G_{\ell_1m_1}V^{G\,*}_{\ell_2m_2}\rangle \langle V^{G\,*}_{\ell_2'm_2'}V^G_{\ell_1'm_1'}\rangle\,.\nonumber
\ee
Note that the first term in the second line of Eq.~\eqref{eq:four_point} is proportional to $\langle p_{LM}^{*}\rangle\langle p_{L'M'}^{*}\rangle$ and only contributes to the monopole\footnote{This is due to the relation
\be
  \sum\limits_{m}(-1)^{-m}\begin{pmatrix}
   \ell & \ell & L \\
   m & -m & 0 
  \end{pmatrix}=(-1)^\ell\sqrt{2\ell-1}\delta_{L0}
\ee
which we assume to be correct for the reality condition we have chosen. Traditionally there is a $(-1)^{m}$ term that is involved that one must apply to both coefficients and the spherical harmonics functions. In cosmology literature the  definition we used is more common.}, i.e. vanishes for modes $\{L,L'\}\geq1$. For {second and
third} terms of Eq.~\eqref{eq:four_point}, which are identical, the identities
\be
  \sum\limits_{m m'}
  \begin{pmatrix}
   \ell & \ell' & L \\
   m & m' & M 
  \end{pmatrix}
  \begin{pmatrix}
   \ell & \ell' & L' \\
   m & m' & M' 
  \end{pmatrix}=(2L+1)^{-1}\delta_{LL'}\delta_{MM'}\,,\non
\ee
\be
\begin{split}
  \begin{pmatrix}
   \ell_1 & \ell_2 & \ell_3 \\
   s_1 & s_2 & s_3 
  \end{pmatrix}
  &\begin{pmatrix}
   \ell_1 & \ell_2 & \ell_3 \\
   s_1' & s_2' & s_3' 
  \end{pmatrix}=\!\frac{1}{2}\!\int_{-1}^{1}\! \dd(\cos{\theta}) d^{\ell_1}_{s_1s_1'}(\theta)d^{\ell_2}_{s_2s_2'}(\theta)d^{\ell_3}_{s_3s_3'}(\theta)\,,
\end{split}\non
\ee
{ where $d$ is the Wigner-$d$ matrix~\citep{Wigner1931GruppentheorieUI}}, and $d^\ell_{s_1s_1'}=d^\ell_{-s_1-s_1'}$ are useful. Using these, we find the correlation function $\langle p_{L M} p_{L'M'}^{*}\rangle=\delta_{LL'}\delta_{MM'}C^{pp}_L$ as
\be\label{eq:kinquad_signal}
C_\ell^{pp}=4\pi f^2(x)\int_{-1}^{1}\dd(\cos\theta)\xi^{VV}(\theta)\xi^{VV}(\theta)d^{\ell}_{22}(\theta)\,,\non
\ee
where we defined
\be
\xi^{VV}(\theta,\chi) \equiv \sum\limits_\ell\frac{(2\ell+1)}{4\pi}C_\ell^{VV}d_{11}^{\ell}(\theta)\,,
\ee
and
\be\label{eq:velo_signal}
C_\ell^{VV}=4\pi\ell(\ell+1)\int\frac{k^2\dd k}{2\pi^2}P_v(k,\chi)\left[\frac{j_\ell(k\chi)}{k\chi}\right]^2\,,
\ee
where $P_v(k)$ is the dimensionless power spectrum of the three-dimensional velocity. Having calculated the kinematic quadrupole signal, we turn to estimating its variance in the presence of primary CMB polarization and noise. 

\subsection{Contribution to the CMB}

The $E$- and $B$- mode CMB polarization satisfy 
\be\label{eq:EandB1}
E_{\ell m}&=&\frac{1}{2}(\,_{+2}a_{\ell m}+\,_{-2}a_{\ell m})\\
\label{eq:EandB2}
B_{\ell m}&=&\frac{1}{2i}(\,_{+2}a_{\ell m}-\,_{-2}a_{\ell m})
\ee

where
\be
\,_{\pm2}a_{\ell m}=\int\dd^2\bn\,(Q\pm iU)(\bn)\,_{\pm2}Y^*_{\ell m}(\bn)\,.
\ee
Next, we expand the optical depth with an homogeneous and anisotropic part $\dot{\tau}(\chi\bn)=\sigma_T a(\chi) \bar{n}_e(\chi)[1+\delta_e(\chi\bn)]$ where $\bar{n}_e(\chi)$ is the mean electron density at comoving distance $\chi$. The homogeneous part gives 
\be
\,_{\pm2}a_{\ell m}&=&f(x) \int\dd\chi \dot{\bar{\tau}}(\chi)\sum_{\ell' m'}\sum_{\ell'' m''}V^G_{\ell' m'}V^G_{\ell'' m''}\,_{\pm1} Y_{\ell' m'}(\bn)\,_{\pm1} Y_{\ell'' m''}(\bn)\,_{\pm2}Y_{\ell m}^*(\bn)\\
&=&f(x) \int\dd\chi \dot{\bar{\tau}}(\chi)\sum_{\ell' m'}\sum_{\ell'' m''}V^G_{\ell' m'}V^G_{\ell''m''}\,_{\pm}\mathcal{I}_{\ell'\ell''\ell}^{m'm'm}\,.
\ee
Using Eqs.~\eqref{eq:EandB1}~and~\eqref{eq:EandB2}, the kpSZ induced CMB polarization satisfies
\be
E_{\ell m}^{(0)}&=&f(x) \int\dd\chi \dot{\bar{\tau}}(\chi)\sum_{\ell' m'}\sum_{\ell'' m''}\frac{1}{2}[1+(-1)^{\ell'+\ell''+\ell}]V^G_{\ell' m'}V^G_{\ell''m''}
\,_{\pm}\mathcal{I}_{\ell'\ell''\ell}^{m'm''m}\,,
\\
B_{\ell m}^{(0)}&=&f(x) \int\dd\chi \dot{\bar{\tau}}(\chi)\sum_{\ell' m'}\sum_{\ell'' m''}\frac{1}{2}[1-(-1)^{\ell'+\ell''+\ell}]V^G_{\ell' m'}V^G_{\ell''m''}
\,_{\pm}\mathcal{I}_{\ell'\ell''\ell}^{m'm''m}\,.
\ee
where $(0)$ superscript denotes the polarization induced by the homogeneous optical depth. The CMB spectra can be calculated as 
\be\label{eq:EE0}
C_{\ell}^{EE,\,(0)}&=&f(x)f(x')\int_{-1}^{1}\dd\cos(\theta)\sum_{\ell'\ell''}[1+(-1)^{\ell'+\ell''+\ell}]f^{VV}_{\ell'\ell''}d_{11}^{\ell'}(\theta)d_{11}^{\ell''}(\theta)d_{22}^{\ell}(\theta)\\ 
\label{eq:BB0}
C_{\ell}^{BB,\,(0)}&=&f(x)f(x')\int_{-1}^{1}\dd\cos(\theta)\sum_{\ell'\ell''}[1-(-1)^{\ell'+\ell''+\ell}]f^{VV}_{\ell'\ell''}d_{11}^{\ell'}(\theta)d_{11}^{\ell''}(\theta)d_{22}^{\ell}(\theta)
\ee
where
\be
f^{VV}_{\ell\ell'}=\frac{(2\ell+1)(2\ell'+1)}{4\pi}\int\dd\chi\dot{\bar{\tau}}(\chi)\int\dd\chi' \dot{\bar{\tau}}(\chi')C_\ell^{VV}(\chi,\chi')C_{\ell'}^{VV}(\chi,\chi')
\ee

\begin{figure}[t!]
    \includegraphics[width=0.95\columnwidth]{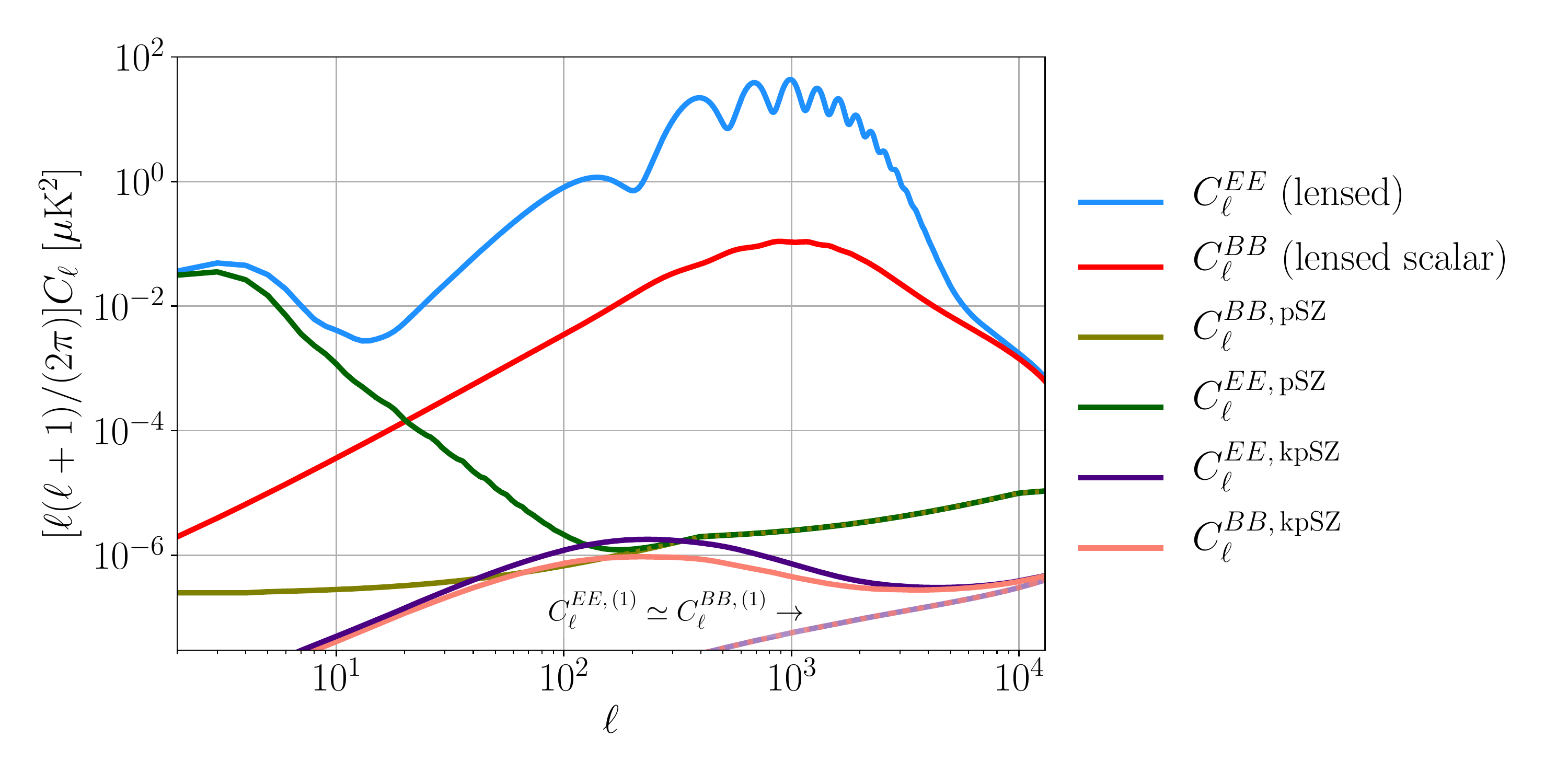}
    \caption{The power spectra of the CMB linear-polarization signals. The blue (red) curve is the lensed primary CMB $E$-mode ($B$-mode) polarization. The green and olive-color lines correspond to the polarized Sunyaev Zel'dovich (pSZ) effect due to the primary quadrupole on the $E$- and $B$-mode fluctuations, respectively. Note that the E-mode pSZ at low-$\ell$ is simply the `reionization bump'. The purple solid and pink dotted lines correspond to the kinetically-induced Sunyaev Zel'dovich polarization (kpSZ) signal on the $E$- and $B$-mode fluctuations due to peculiar transverse velocities of the electrons respectively.}
    \label{fig:cmb_signals}
\end{figure}

\begin{figure}[t!]
    \includegraphics[width=0.95\columnwidth]{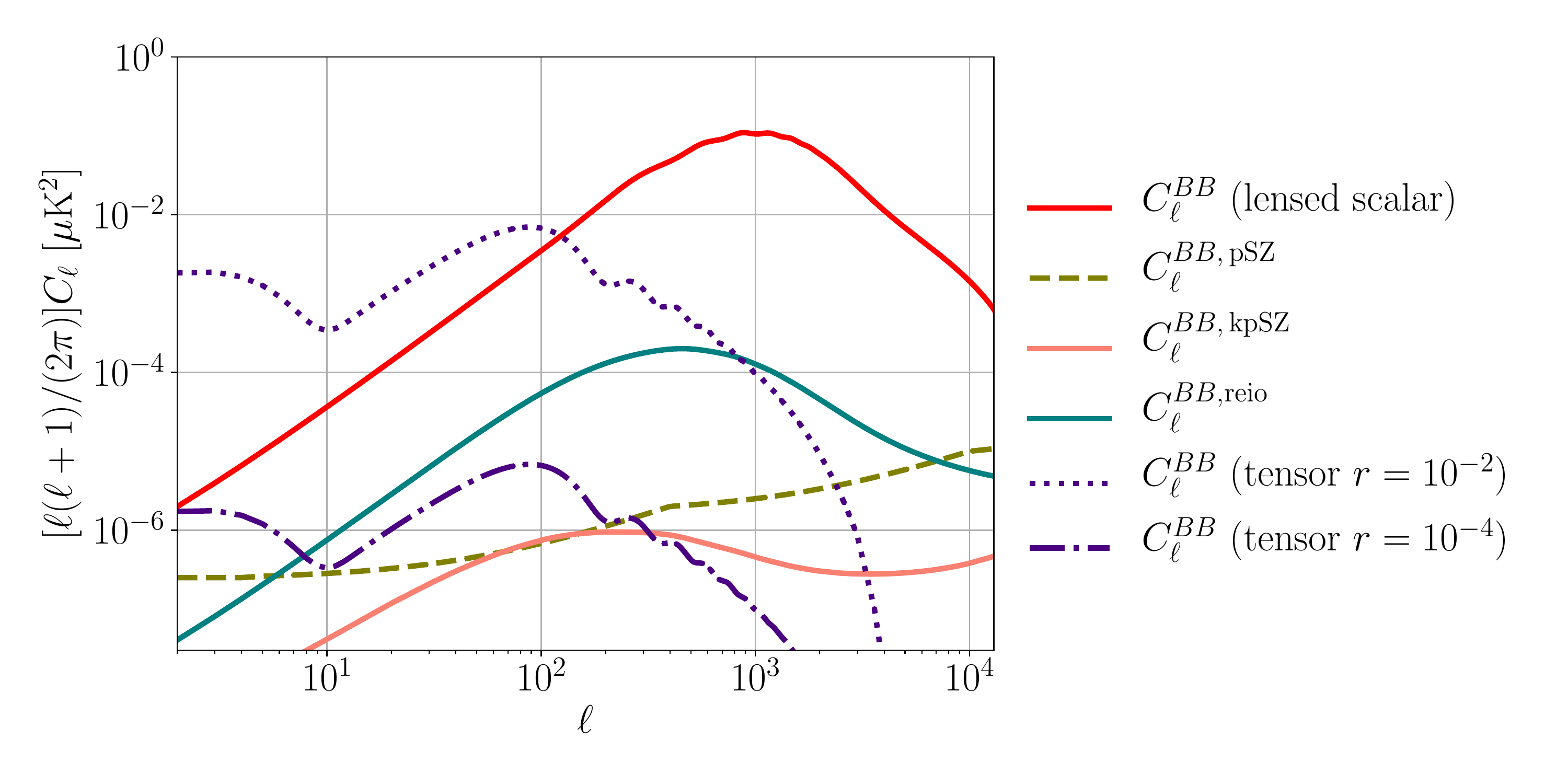}
    \vspace*{-0.8cm}
    \caption{Contributions to the $B$-mode polarization power spectra of the CMB. The red curve is the lensed primary CMB $B$-mode polarization. The olive-color dashed line correspond to the $B$-mode contribution from the polarized Sunyaev Zel'dovich (pSZ) effect due to the primary quadrupole. The purple dotted and dashed-dotted lines tensor contributions to the $B$-mode spectra for the tensor-to-scalar ratio satisfying $r=\{10^{-2},10^{-4}$\}, respectively. The teal solid line is the contribution due to patchy reionization. The salmon solid line correspond to the $B$-mode contribution from the kpSZ signal due to peculiar transverse velocities of electrons.}
    \label{fig:cmb_signals_BB}
\end{figure}

On small scales the fluctuations of the electron density become large. Assuming cross correlations between $\tau(\chi\bn)$ and the velocity are subdominant compared to auto-correlations of $\tau(\chi\bn)$ and $V^G(\chi\bn)$, we can calculate the polarization signals as
\be
E_{\ell m}^{(1)}\simeq-{\sigma_T}\int\dd\chi_e a_e \bar{n}_e(\chi_e)\sum\limits_{\ell' m'}\sum\limits_{\ell''m''}&&\frac{1}{2}\left(1+(-1)^{\ell'+\ell''+\ell}\right)\,_{\pm}\mathcal{J}_{\ell'\ell''\ell}^{m'm''m}\delta_{\ell'' m''}(\chi_e)p_{\ell' m'}(\chi_e)\,,\,\,
\ee
and
\be
B_{\ell m}^{(1)}\simeq-{\sigma_T}\int\dd\chi_e a_e \bar{n}_e(\chi_e)\sum\limits_{\ell' m'}\sum\limits_{\ell''m''}&&\frac{1}{2}\left(1-(-1)^{\ell'+\ell''+\ell}\right)\,_{\pm}\mathcal{J}_{\ell'\ell''\ell}^{m'm''m}\delta_{\ell'' m''}(\chi_e)p_{\ell' m'}(\chi_e)\,,\,\,
\ee
where  
\be
\,_{\pm}\mathcal{J}_{\ell'\ell''\ell}^{m'm''m}=\sqrt{\frac{(2\ell'+1)(2\ell''+1)(2\ell+1)}{4\pi}}\begin{pmatrix}
  \ell & \ell' & \ell'' \\
  m' & m'' & -m 
  \end{pmatrix}
  \begin{pmatrix}
  \ell & \ell' & \ell'' \\
  -2 & 2 & 0
  \end{pmatrix}\,.
\ee
The total $E$- and $B$-mode power-spectra for the kpSZ signal 
\be
C_{\ell}^{EE,\,\rm kpSZ}=C_{\ell}^{EE\,(0)}+C_{\ell}^{EE\,(1)}\,,\\
C_{\ell}^{BB,\,\rm kpSZ}=C_{\ell}^{BB\,(0)}+C_{\ell}^{BB\,(1)}\,,
\ee
are shown in Fig.~\ref{fig:cmb_signals} together with the $E$- and $B$-mode pSZ signals from the primordial quadrupole and the lensed primary spectra.\footnote{Note that unlike pSZ, the kpSZ $B$-modes are sourced both at linear and non-linear orders, and are dominated by the linear contributions.} The kpSZ effect remains sub-dominant compared to the $E$-mode spectrum induced by the pSZ effect on most scales, and is lower than the contribution due to patchy reionization by over an order of magnitude on all scales. These effects can be a limiting factor in $B$-mode searches for small $r$ as shown in Fig.~\ref{fig:cmb_signals_BB},  depending on the effectiveness of CMB delensing, see~e.g.~Refs.~\citep{Knox:2002pe,Kesden:2002ku,Seljak:2003pn,Smith:2010gu,Abazajian:2016yjj,Abazajian:2019eic,CMB-S4:2020lpa,Carron:2017vfg,Planck:2018lbu,Green:2016cjr,Hotinli:2021umk}. {Throughout this work, we set the cosmological parameters $\{\Omega_c h^2,\Omega_b h^2,\theta_s,\tau,A_s,n_s\}$ to $\{0.1197,0.022,0.0104,0.06,2.196\!\times\!10^{-9},0.9655\}$, respectively, where $\Omega_c h^2$ ($\Omega_b h^2$) is the physical cold dark matter (baryon) density, $\theta_s$ is the angle subtended by acoustic scale, $\tau$ is the Thomson optical depth to recombination, $A_s$ is the primordial scalar fluctuation amplitude and $n_s$ is the primordial scalar fluctuation slope.}

\section{Detecting the signal}\label{sec:detection}

\begin{table}[t]
\begin{tabular}{|l|c|c|c|c|}
\hline
& \multicolumn{2}{c|}{Beam FWHM} & \multicolumn{2}{c|}{Noise RMS} \\
& \multicolumn{2}{c|}{} & \multicolumn{2}{c|}{($\mu$K-arcmin)} \\ 
\hline
& \multicolumn{1}{c|}{S4} & \multicolumn{1}{c|}{HD} &
\multicolumn{1}{c|}{S4} & \multicolumn{1}{c|}{HD} \\ \hline
39 GHz                  & $5.1'$                  & $36.3''$                                     & 12.4                     & 3.4                     \\
93 GHz                  & $2.2'$                  & $15.3''$                                      & 2.0                     & 0.6                     \\
145 GHz                 & $1.4'$                  & $10.0''$                                     & 2.0                     & 0.6                     \\
225 GHz                & $1.0'$                  & $6.6''$                                       & 6.9                     & 1.9                     \\
280 GHz                 & $0.9'$                  & $5.4''$                                      & 16.7                    & 4.6                     \\ \hline
\end{tabular}
\caption{{\it Inputs to ILC noise for the baseline CMB configurations:} The beam and temperature noise RMS parameters are chosen to roughly match CMB-S4 and CMB-HD. We model the CMB noise as shown in Eq.~\eqref{eq:CMB_noise}. In both cases, we account for the degradation due to Earth's atmosphere by defining the CMB noise choose $\ell_{\rm knee}=100$ and $\alpha_{\rm knee}=-3$. The polarization noise satisfy $\Delta_E=\Delta_B=\sqrt{2}\Delta_T$.}
\label{tab:beamnoise}
\end{table}

\begin{table}[b]
\begin{tabular}{|l|c|c|c|c|c|c|c|}
\hline
Frequency (GHz) & 20 & 27 & 39 & 93 & 145 & 225 & 278 \\
Beam FWHM (arcmin)&11.0&8.4&5.8&2.5&1.6&1.1&1.0 \\
\hline
noise $T$($\mu K$-arcmin) & 9.31&4.60&2.94&0.45&0.41&1.29&3.07 \\
\hline
noise E ($\mu K$-arcmin)&13.16&6.50&4.15&0.63&0.59&1.83&4.34 \\
\hline
noise $B$ ($\mu K$-arcmin)&13.16&6.50&4.15&0.63&0.59&1.83&4.34 \\ 
\hline
\end{tabular}
\caption{{\it Inputs to ILC noise for the CMB-S4 V3R0-25 configuration}. For CMB-HD, we take the same configuration with the improved CMB-HD beam given on Table~I. We model the CMB noise as shown in Eq.~\eqref{eq:CMB_noise} and take $\ell_{\rm knee}=100$ and $\alpha_{\rm knee}=-3$ for all frequency bins and observables.}
\label{tab:beamnoise2}
\end{table}

\begin{figure}[t!]
    \includegraphics[width=1.0\columnwidth]{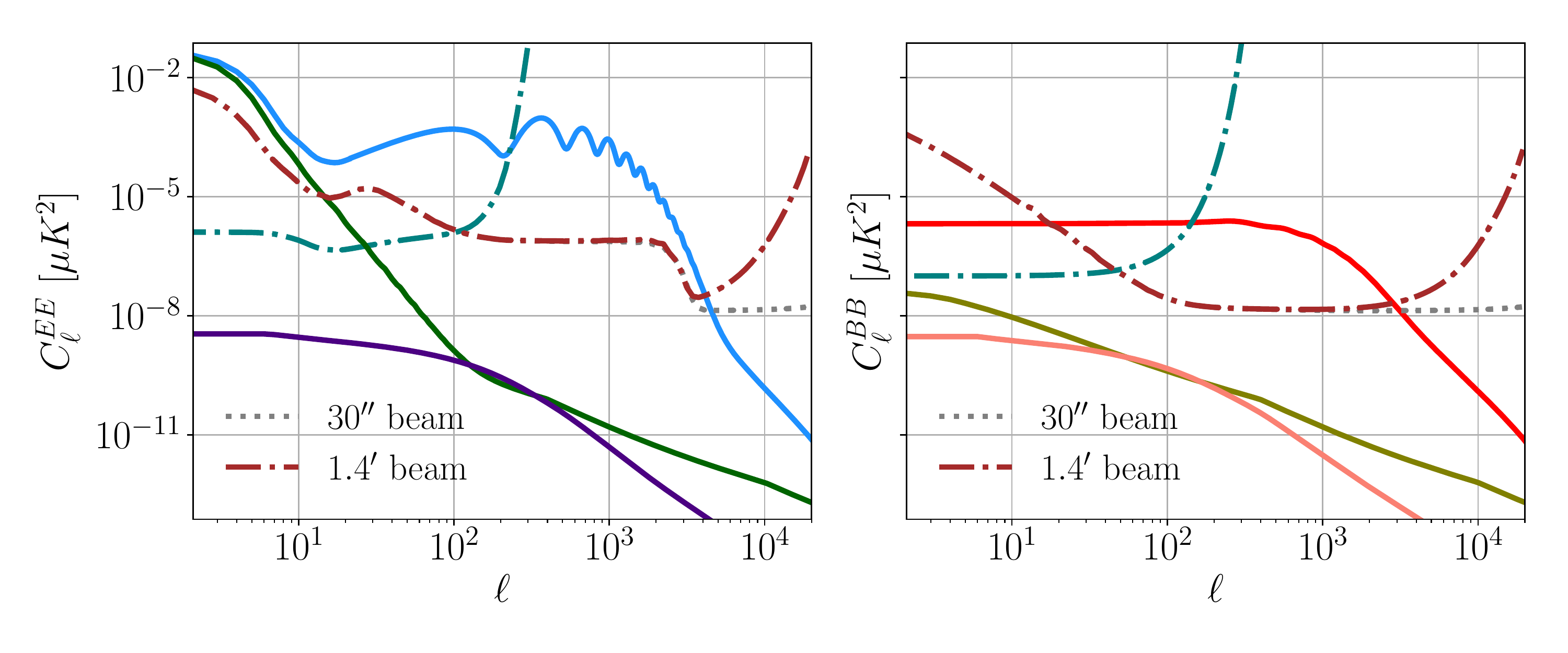}\\
    \vspace{-0.5cm}\hspace{1.7cm}\includegraphics[width=0.57\columnwidth]{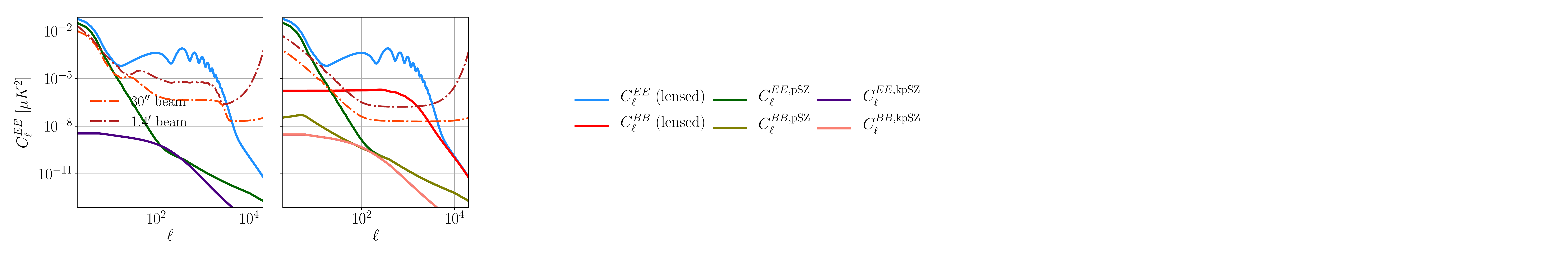}
    \caption{\textit{The kpSZ signal and the background.} The CMB $E$- and $B$-mode spectra are shown with blue and red solid lines respectively. The former contains the primary pSZ contribution the $E$-mode spectrum, which is shown with the green solid line. The $B$-mode contribution form the pSZ effect is shown with the olive-color line on the right plot. The ILC-cleaned foregrounds for the kpSZ measurement (the primary CMB and the pSZ signals) are shown with the dot-dashed brown and dotted gray lines for CMB-S4  and CMB-HD like surveys, respectively, using the V3R0-25 configuration. The teal-colored dot-dashed lines correspond to the ILC-cleaned noise curves anticipated from the PIXIE experiment. The kpSZ signal on the $E$- and $B$-mode spectra are shown with the purple and pink solid lines, respectively. The ILC-cleaning is performed by appropriating the frequency dependence of the kpSZ signal to the CMB primary, the pSZ signal, and the noise (both white and pink) as described in the text. }
    \label{fig:cleaned_cmbE}
\end{figure}

\subsection{ILC-cleaned kpSZ maps}\label{sec:ILC-cleaning}

The frequency dependence of the kpSZ signal provides a window into removing some of the lensed primary CMB and the pSZ signals -- both of which are black-body -- that are dominant in comparison to the kpSZ. The multi-frequency information can be used to clean  contributions to the temperature maps other than the kpSZ signal using a standard harmonic-space internal linear combination (ILC)
procedure. We write the covariance between the de-beamed CMB at different frequencies as a matrix satisfying 
\be
\boldsymbol{C}_\ell=\tilde{C}_\ell^{EE,\,\rm kpSZ}\boldsymbol{e}\boldsymbol{e}^\dagger+\boldsymbol{g}^{-1}_{\rm kpSZ}{C}_\ell^{FG}+(\boldsymbol{B}^{-1}\boldsymbol{g}^{-1}_{\rm kpSZ}\boldsymbol{N})_\ell
\ee
where $(\boldsymbol{g}^{-1}_{\rm kpSZ})_{ij}=1/[g(x_i)g(x_j)]$ where $g(x_i)$ is the frequency-dependent coefficient of the kpSZ signal and $\tilde{C}^{EE,\,\rm kpSZ}_\ell$ is the frequency-independent component of the kpSZ signal introduced in Eq.~\eqref{eq:quad_source_kin}, i.e. $\tilde{C}^{EE,\,\rm kpSZ}_\ell={C}^{EE,\,\rm kpSZ}_\ell/[g(x)]^2$. Here, ${C}_\ell^{FG}$ consists  of the non-kpSZ contribution (foregrounds for kpSZ) to the CMB (the primary lensed blackbody CMB and pSZ signals) and $(\boldsymbol{B}^{-1}\boldsymbol{N})_\ell$ is the de-beamed instrumental noise covariance (assumed diagonal). 

{For the contributions to the signal in the millimeter wavelength, we consider the black-body late-time kSZ, clustered CIB, Poisson CIB and tSZ foregrounds following~Ref.~\citep{Madhavacheril:2017onh}. We omit including the correlation between CIB and tSZ. We include the radio sources using the flux-limit-dependent radio-source power model from Ref.~\citep{Lagache19} in the 39, 93 and 145 GHz channels. We assume flux limits of 10, 7 and 10 mJy (2, 1 and 1 mJy) for CMB-S4 (CMB-HD), respectively in those channels. Finally, we include the kSZ signal from  reionization following Ref.~\citep{Park:2013mv} and the lensed CMB black-body contribution using CAMB~\citep{CAMB}.}

Following
the ILC method in harmonic space, we estimate the ILC-cleaned signal as 
\be
\hat{E}_{\ell m}=\boldsymbol{w}^\dagger_{\ell m}\boldsymbol{E}_{\ell m}
\ee
where $\boldsymbol{w}_{\ell m}$ are weights that minimize the variance of the resulting signal satisfying
\be
\boldsymbol{w}_{\ell m}=\frac{
(\boldsymbol{C}_\ell)^{-1}
\boldsymbol{e}}
{\boldsymbol{e}^\dagger(\boldsymbol{C})^{-1}\boldsymbol{e}}\,,
\ee
and the ensemble averaged $E$-mode power spectrum (same for the $B$-mode) of the cleaned map is
\be
C_\ell^{EE,\,\rm cleaned}=\tilde{C}_\ell^{EE,\,\rm kpSZ}+\boldsymbol{w}_\ell^\dagger[\boldsymbol{g}^{-1}_{\rm kpSZ}{C}_\ell^{FG}+(\boldsymbol{B}^{-1}\boldsymbol{g}^{-1}_{\rm kpSZ}\boldsymbol{N})_\ell]\boldsymbol{w}_\ell\,.
\ee

We demonstrate the cleaned $E$-mode polarization spectra in Fig.~\ref{fig:cleaned_cmbE} for CMB-S4 and CMB-HD like survey specifications (dot-dashed lines). We find that although the primary CMB and pSZ signals are removed significantly from the cleaned $E$- and $B$-maps, some residual from these dominant foregrounds still remain and the CMB background on the kpSZ measurement is still $\sim\mathcal{O}(10^{2})$ larger than the experimental white-noise. As the number of frequency bins plays an important role in reducing the dominant blackbody contribution, the proposed PIXIE experiment~\citep{2016SPIE.9904E..0WK,2014SPIE.9143E..1EK} with 15GHz wide frequency bins within the range 30GHz-3THz can potentially provide much more significant reduction. Using the experimental specifications of PIXIE, we find the blackbody signal can be removed almost perfectly from the CMB allowing PIXIE to provide detector-noise limited measurement of the kpSZ signal up to scales $\ell\sim100$ allowed by its 2.6 degree beam. In what follows we provide forecasts for the `direct-detection' prospects for the kpSZ effect using CMB-S4, CMB-HD and PIXIE experiments. In addition to these upcoming LiteBIRD~\citep{Matsumura:2013aja,Hazumi:2019lys} and proposed ESA Voyage 2050~\citep{Basu:2019rzm,Delabrouille:2019thj} surveys can also potentially improve the prospects of detecting this effect directly on large scales by cross-correlating with velocity templates.

\subsection{Direct detection}\label{sec:direct_detection}

\begin{figure}[t!]
    \includegraphics[width=0.7\columnwidth]{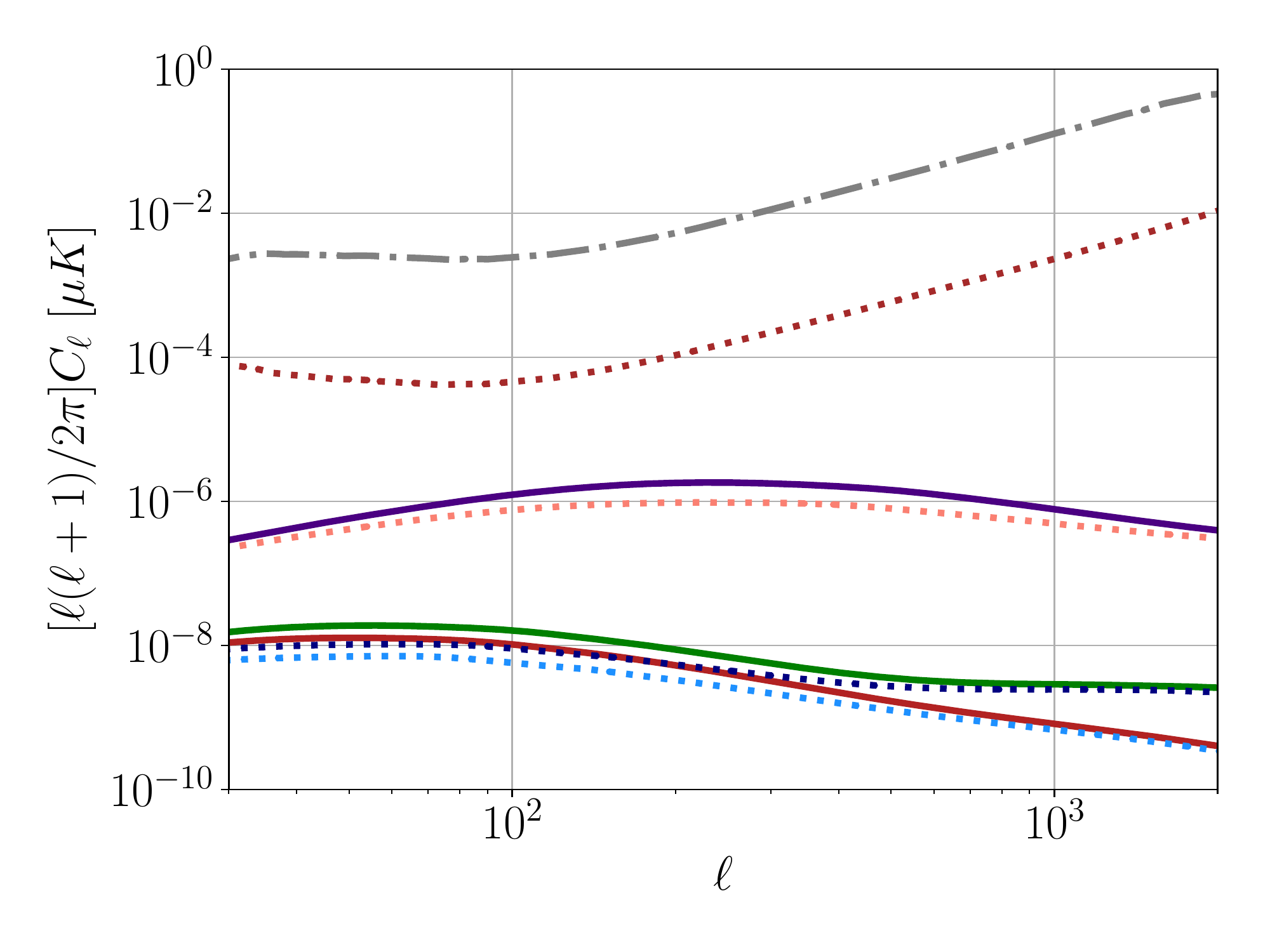}\\
    \vspace*{-0.5cm}
    \includegraphics[width=0.7\columnwidth]{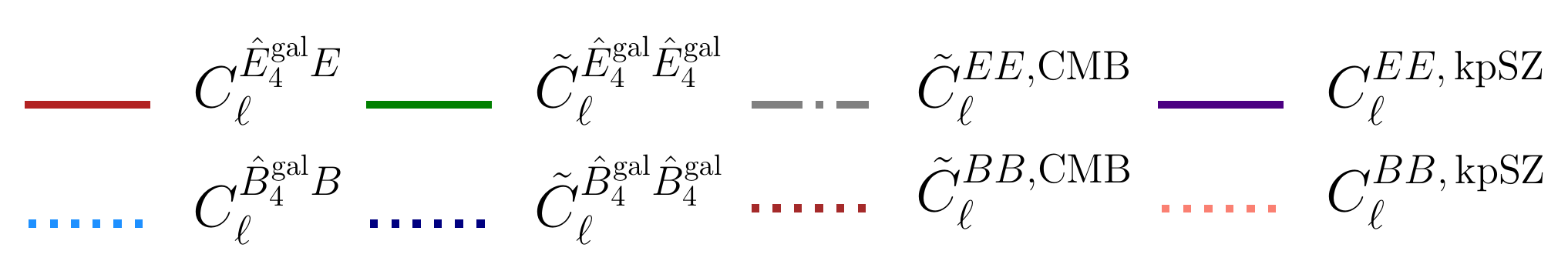}
    \vspace*{-0.3cm}
    \caption{\textit{The kpSZ signal and noise.} The redshift-binned polarization signals reconstructed from a galaxy survey shown together with the cross-correlation signal between the CMB. The two lines show the anticipated CMB variance after ILC-cleaning as discussed in the text. The third and fourth (identical) lines from the top are the total kpSZ signal in the CMB maps. We find the cross-correlation signal between the polarization maps reconstructed from a galaxy survey at a given redshift bin and the CMB is comparable to the polarization signal autocorrelation reconstructed from a galaxy survey on scales of interest for detection. Here, the galaxy survey is taken to match the experimental specifications of VRO. The ILC-cleaned CMB lines correspond to the noise estimates from the ultra-deep V3R0-25 CMB-S4 configuration. We show the signal from the fourth galaxy redshift bin and we separate the redshift range $z\in[0.2,6]$ into 32 redshift bins.}
    \label{fig:signal_direct}
\end{figure}
\begin{figure}[t!]
    \includegraphics[width=0.94\columnwidth]{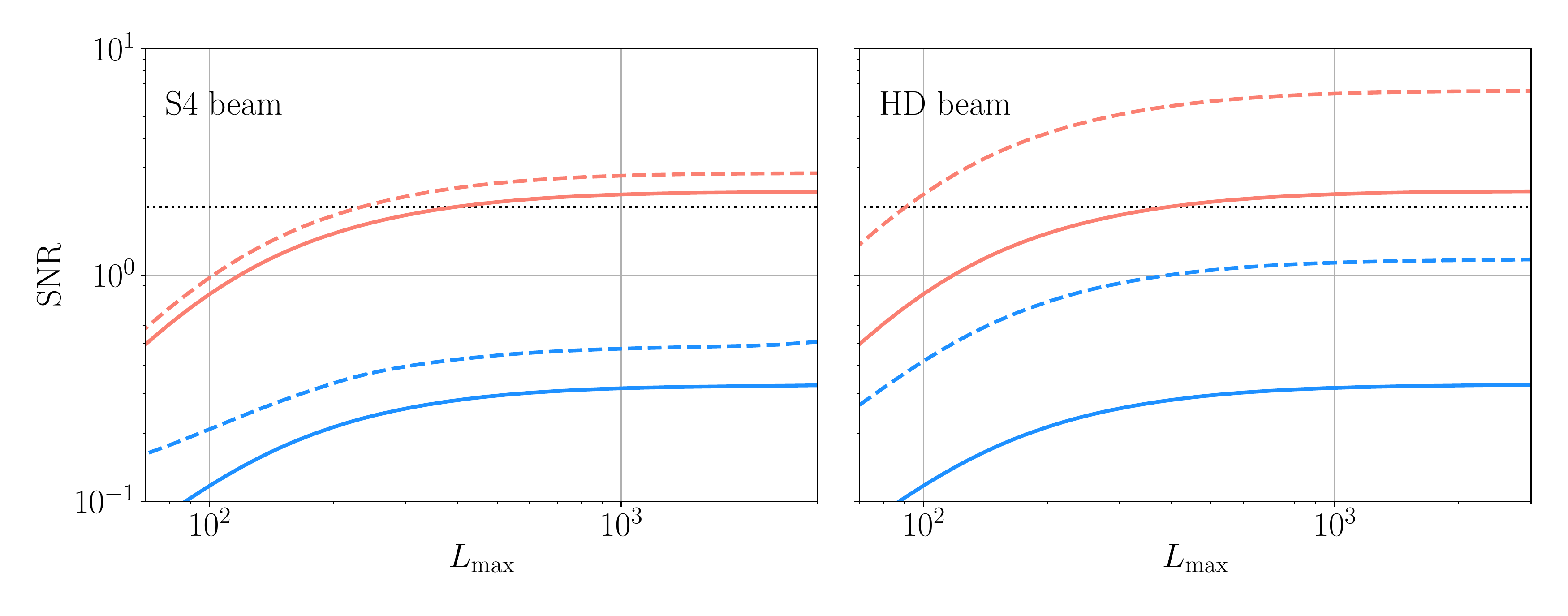}
    \includegraphics[width=0.55\columnwidth]{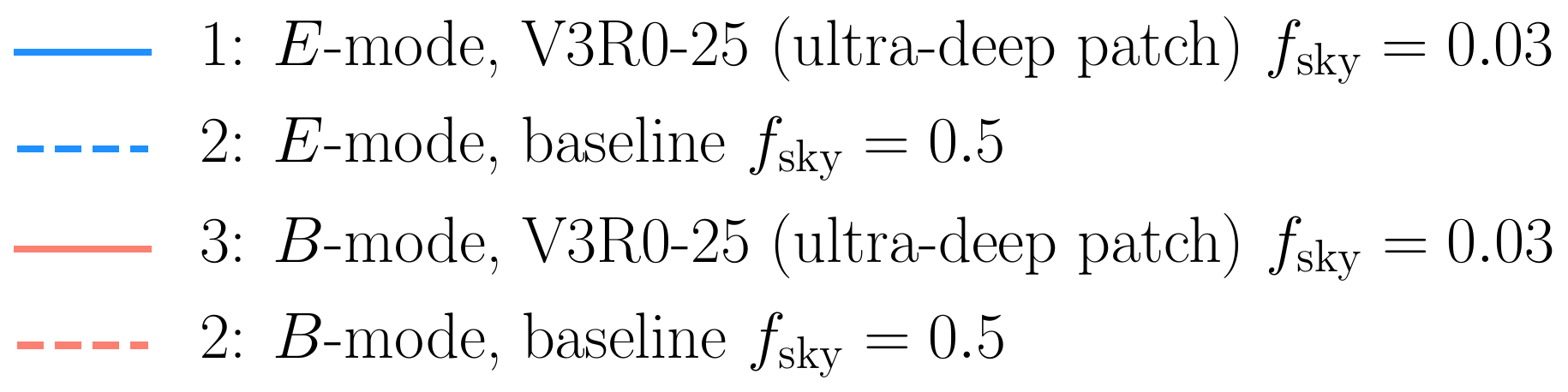}
    \vspace*{-0.4cm}
    \caption{\textit{Detection prospects of the kpSZ signal from cross-correlating ILC-cleaned kpSZ maps with velocity templates.} We use beams matching the experimental specifications of the CMB-S4 and the CMB-HD surveys. The kinetic quadrupole is reconstructed using Vera Rubin Observatory. We show the SNR as a function of the maximum reconstructed velocity multipole $L_{\rm max}$ from the galaxy survey.}
    \label{fig:snr_direct}
\end{figure}

\begin{figure}[t!]
    \includegraphics[width=0.5\columnwidth]{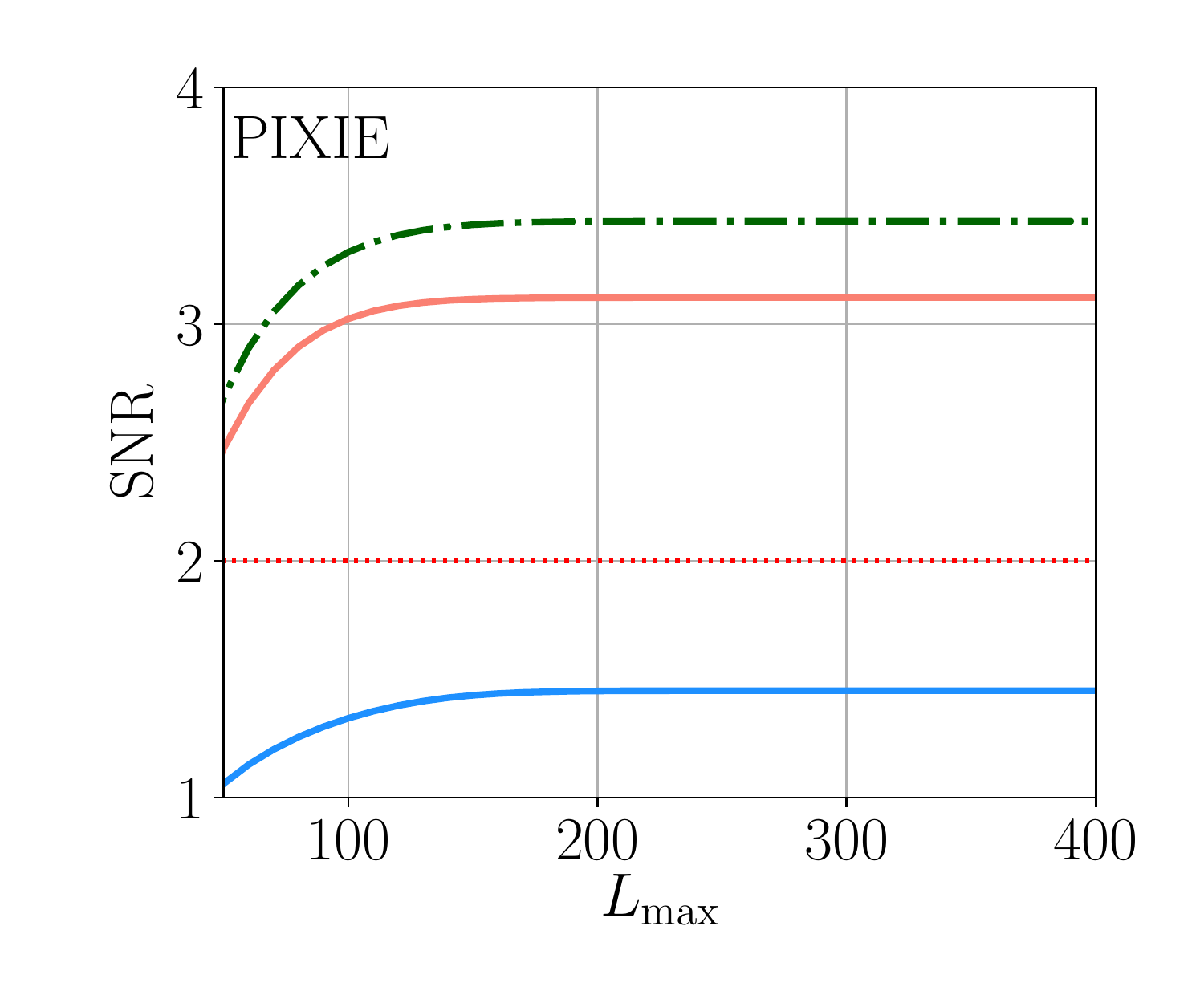}\\
    \vspace*{-0.3cm}
    \includegraphics[width=0.35\columnwidth]{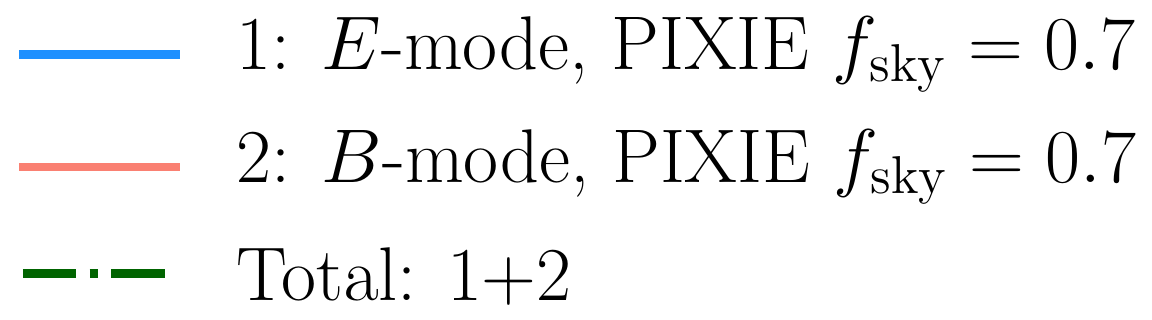}
    \vspace*{-0.3cm}
    \caption{\textit{Detection prospects of the kpSZ signal from cross-correlating ILC-cleaned kpSZ maps with velocity templates.} We use CMB experimental specifications matching the proposed PIXIE survey. Figure otherwise similar to Fig.~\ref{fig:snr_direct}}
    \label{fig:snr_direct_PIXIE}
\end{figure}

One of the most promising directions for a first detection of the kpSZ effect is via cross-correlating the degree-scale kinetic quadrupole reconstructed from a galaxy survey with the $E$- and $B$-mode polarizations measured from the CMB. Although the lensed primary CMB is orders of magnitude larger than the kpSZ signal, the frequency dependence of the latter can be used to increase the prospects to detect this signal as we discussed in the previous section. 

The cross-correlation of the kpSZ $E$- and $B$-mode fluctuations in the CMB and the reconstructed field satisfies the same equalities in Eqs.~\eqref{eq:EE0}~and~\eqref{eq:BB0} with 
\be
f^{VV}_{\ell\ell'}\rightarrow f^{V\hat{V}_{\rm gal}^{(z_i)}}_{\ell\ell'}=\frac{(2\ell+1)(2\ell'+1)}{4\pi}\int\dd\chi\dot{\bar{\tau}}(\chi)\int\dd\chi' {W}_i(\chi')\dot{\bar{\tau}}(\chi')C_\ell^{V\hat{V}_{\rm gal}}(\chi,\chi')C_{\ell'}^{V\hat{V}_{\rm gal}}(\chi,\chi')\,,\non
\ee
and
\be
C_{\ell}^{V\hat{V}_{\rm gal}}=\frac{[C_\ell^{Vg}(\chi,\chi')]^2}{C_\ell^{gg}(\chi')+N_\ell^{gg}(\chi')}\,,
\ee
where $C_\ell^{Vg}(\chi,\chi')$ is the velocity-galaxy cross-correlation, $W_i(\chi)$ is the window function set equal to zero outside the redshift range of the $i$th bin, and $C_\ell^{gg}(\chi')+N_\ell^{gg}(\chi')$ is the observed galaxy auto-spectrum at comoving distance $\chi'$. We define the the galaxy-galaxy power spectrum computed using the Halo Model following Refs.~\citep{Smith:2018bpn,Munchmeyer:2018eey} and use the ReCCO simulation software~\citep{Cayuso:2021ljq} for calculating the galaxy-galaxy, galaxy-velocity and velocity-velocity spectra. We model the photometric redshift errors $\sigma_z=\sigma_0(1+z)$ with $\sigma_0=0.3$. We assume shot noise is uncorrelated between redshift bins, and compute the number density per bin assuming the galaxy number density per square arcmin satisfies
\be
n(z)=\frac{n_g}{2z_0}\left(\frac{z}{z_0}\right)^2\exp\left({-\frac{z}{z_0}}\right)
\ee
with $n_g=40/({\rm arcmin})^2$ and $z_0=0.3$.

We calculate the total signal-to-noise ratio (SNR) from measurements of $E$- and $B$-mode CMB polatization by approximating the likelihood as Gaussian 
\be
 {\rm SNR}^2=&\sum_{\ell\ell'; PXY} C_\ell^{\hat{P}^{\rm gal}_X P}\textbf{cov}^{-1}\left(\tilde{C}^{\hat{P}^{\rm gal}_X P}_{\ell},\tilde{C}^{\hat{P}^{\rm gal}_Y P }_{\ell'}\right) C_{\ell'}^{\hat{P}^{\rm gal}_Y P}  \, ,
\ee
where $P\in\{E,B\}$ indicates the CMB polarization, the $\{X,Y\}$ indices run over redshift bins. Here, the fields with hats refer to polarization maps reconstructed from the galaxy survey in a given redshift bin, those without a hat refer to the polarization maps measured from the CMB, the tilde refers to spectra including noise, and the covariance is given by  
\be\label{eq:covdef}
 \textbf{cov}\left(\tilde{C}^{\hat{P}^{\rm gal}_X P}_{\ell},\tilde{C}^{\hat{P}^{\rm gal}_Y P}_{\ell'}\right) =  \frac{\delta_{\ell\ell'}}{2\ell+1} f_\mathrm{sky}^{-1} 
 \left(\tilde{C}_{\ell}^{\hat{P}^{\rm gal}_X \hat{P}^{\rm gal}_Y}\tilde{C}_{\ell}^{P P}+\tilde{C}_{\ell}^{\hat{P}^{\rm gal}_X P}\tilde{C}_{\ell}^{\hat{P}^{\rm gal}_Y P}\right)  \, .
\ee
To assess the detectability of the kpSZ effect, we take a null hypothesis a scenario in which there is no signal or noise in the CMB cross with the galaxy-derived polarization ($\tilde{C}_\ell^{\hat{P}^{\rm gal}_X P}=0$) when calculating the covariance matrix. We also take to have noise diagonal in the redshift bins ($N_\ell^{\hat{P}^{\rm gal}_X\hat{P}^{\rm gal}_Y}=\delta_{\hat{X}\hat{Y}}N_\ell^{\hat{P}^{\rm gal}_X}$). Our covariance matrix has the form
\be
\textbf{cov}\left(\tilde{C}^{\hat{P}^{\rm gal}_X P}_{\ell},\tilde{C}^{\hat{P}^{\rm gal}_Y P}_{\ell'}\right)=
\begin{pmatrix}
\tilde{C}_\ell^{\hat{P}^{\rm gal}_1\hat{P}^{\rm gal}_1}\tilde{C}_\ell^{PP} & {C}_\ell^{\hat{P}^{\rm gal}_1\hat{P}^{\rm gal}_2}\tilde{C}_\ell^{PP} & \ldots & {C}_\ell^{\hat{P}^{\rm gal}_1\hat{P}^{\rm gal}_{N_{\rm bins}}}\tilde{C}_\ell^{PP}\\
{C}_\ell^{\hat{P}^{\rm gal}_2\hat{P}^{\rm gal}_1}\tilde{C}_\ell^{PP} & \tilde{C}_\ell^{\hat{P}^{\rm gal}_2\hat{P}^{\rm gal}_2}\tilde{C}_\ell^{PP} & \ldots & {C}_\ell^{\hat{P}^{\rm gal}_2\hat{P}^{\rm gal}_{N_{\rm bins}}}\tilde{C}_\ell^{PP}
\\
\vdots & \vdots&  \ddots & \vdots
\\
{C}_\ell^{\hat{P}^{\rm gal}_{N_{\rm bins}}\!\!\!\hat{P}^{\rm gal}_1}\tilde{C}_\ell^{PP} & {C}_\ell^{\hat{P}^{\rm gal}_{N_{\rm bins}}\!\!\!\hat{P}^{\rm gal}_2}\tilde{C}_\ell^{PP} & \hdots & \tilde{C}_\ell^{\hat{P}^{\rm gal}_{N_{\rm bins}}\!\!\!\hat{P}^{\rm gal}_{N_{\rm bins}}}\tilde{C}_\ell^{PP}
\end{pmatrix}\,.
\ee

We model the CMB noise with the white and pink noise components as 
\be
N_\ell^{PP}=\Delta_P^2\exp\left[\ell(\ell+1)\frac{\theta_{\rm FWHM}^2}{8\log(2)}\right][1+(\ell_{\rm knee}/\ell)^{\alpha_{\rm knee}}]\label{eq:CMB_noise}
\ee
where we set the polarization noise RMS $\Delta_P$, the beam full-width half-maximum (FWHM) resolution, and the $\{\alpha_{\rm knee},\ell_{\rm knee}\}$ parameters which capture the effect of the degradation due to Earth's atmosphere to match upcoming experiments as defined in Tables~I~and~II.

{ We display the kpSZ signal and noise for direct detection in Fig.~\ref{fig:signal_direct}, where the redshift-binned polarization signals reconstructed from a galaxy survey shown together with the cross-correlation signal between the CMB.} We demonstrate the contributions to the total detection SNR from different maximum multipoles included in the analysis on Fig.~\ref{fig:snr_direct} for CMB-S4 and CMB-HD surveys. We find most of the SNR is contributed from scales satisfying $\ell\in[50,200]$ and that the contribution to the SNR is dominated by the $B$-mode polarization maps. Our forecast suggest that both CMB-S4 and CMB-HD may reach SNR values of $\sim2$, and possibly greater for large sky fraction. In Fig.~\ref{fig:snr_direct_PIXIE} we show results from using the PIXIE survey specifications. The large number of frequency bins and the absence of atmospheric foregrounds allow accessing the largest scale polarization modes $\ell\lesssim100$ almost up to the detector noise precision, suggesting a `direct' direction with ${\rm SNR}\gtrsim3$ may be possible. 

\subsection{The kpSZ quadratic-estimator variance}

Detecting the kpSZ from the polarization maps directly on large scales may be difficult due to atmospheric noise, unaccounted foregrounds and effects of masking, for example. An alternative method for measuring large-scale cosmological fluctuations is through using the non-Gaussian information on the small-scale CMB maps, together with maps of the galaxy distribution in tomographic redshift bins. This technique has been shown to be a very promising avenue to cosmological inference for the upcoming cosmology experiments. Here we go over the pSZ quadratic estimator formalism and evaluate the prospects to detect the large-scale kinematic quadrupole through pSZ tomography. 

The minimum variance estimator for the remote quadrupole field can be reconstructed using observed maps of the CMB polarization and the fluctuations of the optical depth $\delta\tau(\bn)=\sigma_T n_e a \delta_e(\bn)$. The estimator is already calculated in Refs.~\citep{Alizadeh:2012vy,Meyers:2017rtf,Deutsch:2017ybc,Deutsch:2017cja,Deutsch:2018umo}. Here, we shortly review the method presented in Ref.~\citep{Meyers:2017rtf}. Note we are using the flat-sky approximation, which we anticipate to agree well with a full-sky estimator that is calculated in Ref.~\citep{Alizadeh:2012vy}, as this is also the case for lensing estimators~\citep{Okamoto:2003zw}. Parameters $\boldsymbol{\ell}$ and $\boldsymbol{L}$ introduced in this section are two-dimensional Fourier wavevectors.   

We wish to construct an estimator for the quadrupole field $\hat{p}$ from observed maps of the CMB polarization $(Q(\bn)\pm iU(\bn))^{\mathrm{obs}}$, and maps of the local optical depth $\delta\tau^\mathrm{ext}(\bn,\eta)$ observed by some external survey, at small scales,
\be
{\,_\pm}\hat{p}(\bn,\eta)\sim\left[(Q(\bn)\pm iU(\bn))^{\mathrm{obs}}\delta\tau(\bn,\eta)^{\mathrm{ext}}\right]_f\, ,
\ee
filtered in order to minimize the variance of the estimator, similar to CMB lensing quadratic estimator~\citep{Hu:2001kj}. The observed CMB polarization will contain contribution that is primordial in its origin $(Q(\bn)\pm iU(\bn))^{\mathrm{prim}}$, present in the absence of free electrons at late times; as well as contributions from Thomson scattering, ${\,_\pm}p^{\mathrm{pSZ}}(\bn,\eta)\delta\tau^{\mathrm{ext}}(\bn,\eta)$ which include polarization sourced by both primordial and kinematic quadrupoles; as well as instrumental noise. 

The remote quadrupole estimators can be found as
\be
&&\hat{p}^E(\bn,\eta)\\
&&=N^{EE}(\vL)\!\!\int\frac{\dd^2\vl}{(2\pi)^2}\delta\tau^{
\rm ext}(\vL-\vl)f^E(\vl,\vL)[E^{\rm obs}(\vl)\cos^2(2(\varphi_{\vL}\!-\!\varphi_{\vl}))+B^{\rm obs}(\vl)\sin^2(2(\varphi_{\vL}\!-\!\varphi_{\vl}))]\,,\nonumber
\ee
and
\be
&&\hat{p}^B(\bn,\eta)\\
&&=N^{BB}(\vL)\!\!\int\frac{\dd^2\vl}{(2\pi)^2}\delta\tau^{
\rm ext}(\vL-\vl)f^B(\vl,\vL)[E^{\rm obs}(\vl)\sin^2(2(\varphi_{\vL}\!-\!\varphi_{\vl}))+B^{\rm obs}(\vl)\cos^2(2(\varphi_{\vL}\!-\!\varphi_{\vl}))]\,,\nonumber
\ee
where we defined 
\be
{p}_\pm(\vl,\eta)=(p^E\mp i p^B)(\vl,\eta)e^{\mp 2i\phi_{\vl}}\,.
\ee
The filters that minimize the variance can be found as 
\be
f^E(\vl,\vL)=[C_\ell^{EE,\rm obs}\cos^2(2(\varphi_{\vL}\!-\!\varphi_{\vl}))+C_\ell^{BB,\rm obs}\sin^2(2(\varphi_{\vL}\!-\!\varphi_{\vl}))]^{-1}
\ee
and
\be
f^B(\vl,\vL)=[C_\ell^{EE,\rm obs}\sin^2(2(\varphi_{\vL}\!-\!\varphi_{\vl}))+C_\ell^{BB,\rm obs}\cos^2(2(\varphi_{\vL}\!-\!\varphi_{\vl}))]^{-1}\,.
\ee

The estimated variances of the reconstructed field of remote quadrupoles on large scales are then~\citep{Meyers:2017rtf},
\begin{equation}
\label{eq:noise_quad}
N^{EE}(\vL)\!=\!\!\Big[\!\int\!\!\frac{\dd^2{\vl}}{(2\pi)^2}\frac{(C_{|{\vl}-{\vL}|}^{\delta\tau\,g})^2}{C_{|{\vl}-{\vL}|}^{gg\,\rm obs}}\left(C^{EE,\mathrm{obs}}_\ell\cos^2\left[2\left(\varphi_{\vL}\!-\!\varphi_{\vl}\right)\right]+\!C^{BB,\mathrm{obs}}_\ell\sin^2\left[2\left(\varphi_{\vL}\!-\!\varphi_{\vl}\right)\right]\right)^{-1}\Big]^{-1},\non
\end{equation}
\begin{equation}
N^{BB}(\vL)\!=\!\!\Big[\!\int\!\!\frac{\dd^2{\vl}}{(2\pi)^2}\frac{(C_{|{\vl}-{\vL}|}^{\delta\tau\,g})^2}{C_{|{\vl}-{\vL}|}^{gg\,\rm obs}}\left(C^{EE,\mathrm{obs}}_\ell\sin^2\left[2\left(\varphi_{\vL}\!-\!\varphi_{\vl}\right)\right]+\!C^{BB,\mathrm{obs}}_\ell\cos^2\left[2\left(\varphi_{\vL}\!-\!\varphi_{\vl}\right)\right]\right)^{-1}\Big]^{-1},\non
\end{equation}
and $N^{EB}(\vL)=0$ due to symmetry. The $C^\mathrm{obs}$ indicates the ILC-cleaned CMB spectrum including the instrumental noise, and $C_{|{\vl}-{\vL}|}^{\delta\tau g}$ is cross-correlation between the optical depth fluctuations and and externally inferred map of galaxies. More details on derivation can be found in~\citep{Meyers:2017rtf}. 

In order to evaluate the detection prospects of the kinetic pSZ effect from tomography, we cross-correlate the large-scale kinetic quadrupole reconstructed from maps of the CMB (and a density tracer) on small scales, with that inferred directly from a galaxy survey. We assume that the latter method provides a precise enough measurement of the large-scale density that we can infer the large-scale transverse velocity up to galaxy shot noise, which should be a reasonable approximation for the high number densities of galaxies expected in the surveys we are considering. We calculate the total signal-to-noise ratio similar to before as
\be
 {\rm SNR}^2=&\sum_{\ell\ell'; p;XYWZ} C_\ell^{\hat{p}^{\rm gal}_X\hat{p}_Y}\textbf{cov}^{-1}\left(\tilde{C}^{\hat{p}^{\rm gal}_X\hat{p}_Y}_{\ell},\tilde{C}^{\hat{p}^{\rm gal}_W\hat{p}_Z}_{\ell'}\right) C_{\ell'}^{\hat{p}^{\rm gal}_W\hat{p}_Z}  \, ,
\ee
where $p\in\{p^E,p^B\}$ are the reconstructed $E$- and $B$-mode quadrupoles, the indices run over redshift bins, the fields with hats refer to the remote-quadrupole field reconstructed from the CMB, those without a hat refer to the quadrupole reconstructed from the galaxy distribution, the tilde refers to spectra including noise, and the covariance is given by  
\be
 \textbf{cov}\left(\tilde{C}^{\hat{p}^{\rm gal}_X\hat{p}_Y}_{\ell},\tilde{C}^{\hat{p}^{\rm gal}_W\hat{p}_Z}_{\ell'}\right) =  \frac{\delta_{\ell\ell'}}{2\ell+1} f_\mathrm{sky}^{-1} 
 \left(\tilde{C}_{\ell}^{\hat{p}^{\rm gal}_X \hat{p}^{\rm gal}_W}\tilde{C}_{\ell}^{\hat{p}_Y\hat{p}_Z}+\tilde{C}_{\ell}^{\hat{p}^{\rm gal}_X\hat{p}_Z}\tilde{C}_{\ell}^{\hat{p}^{\rm gal}_Y\hat{p}_W}\right)  \, .
\ee
To assess the detectability of the kpSZ effect, we take -- similar to before -- a null hypothesis scenario in which there is no signal in the CMB-reconstructed quadrupole, which we also take to have noise diagonal in the redshift bins ($\tilde{C}_\ell^{\hat{X}\hat{Y}}=\delta_{\hat{X}\hat{Y}}N_\ell^{\hat{X}}$), and no signal or noise in the cross with the galaxy-derived quadrupole ($\tilde{C}_\ell^{\hat{X}^{\rm gal}\hat{Y}}=0$) when calculating the covariance matrix. 

\begin{figure}[t!]
    \includegraphics[width=0.6\columnwidth]{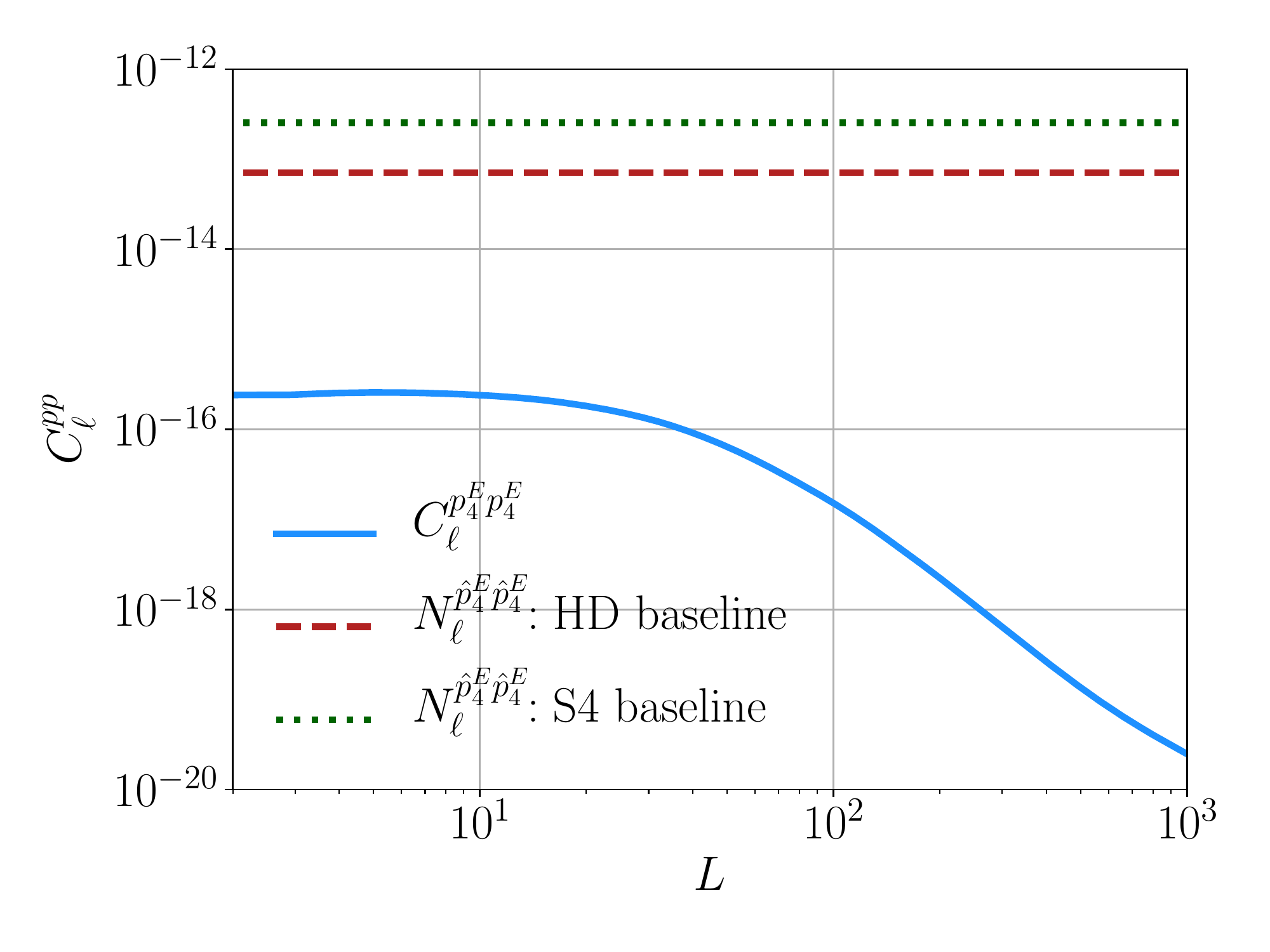}\includegraphics[width=0.3\columnwidth]{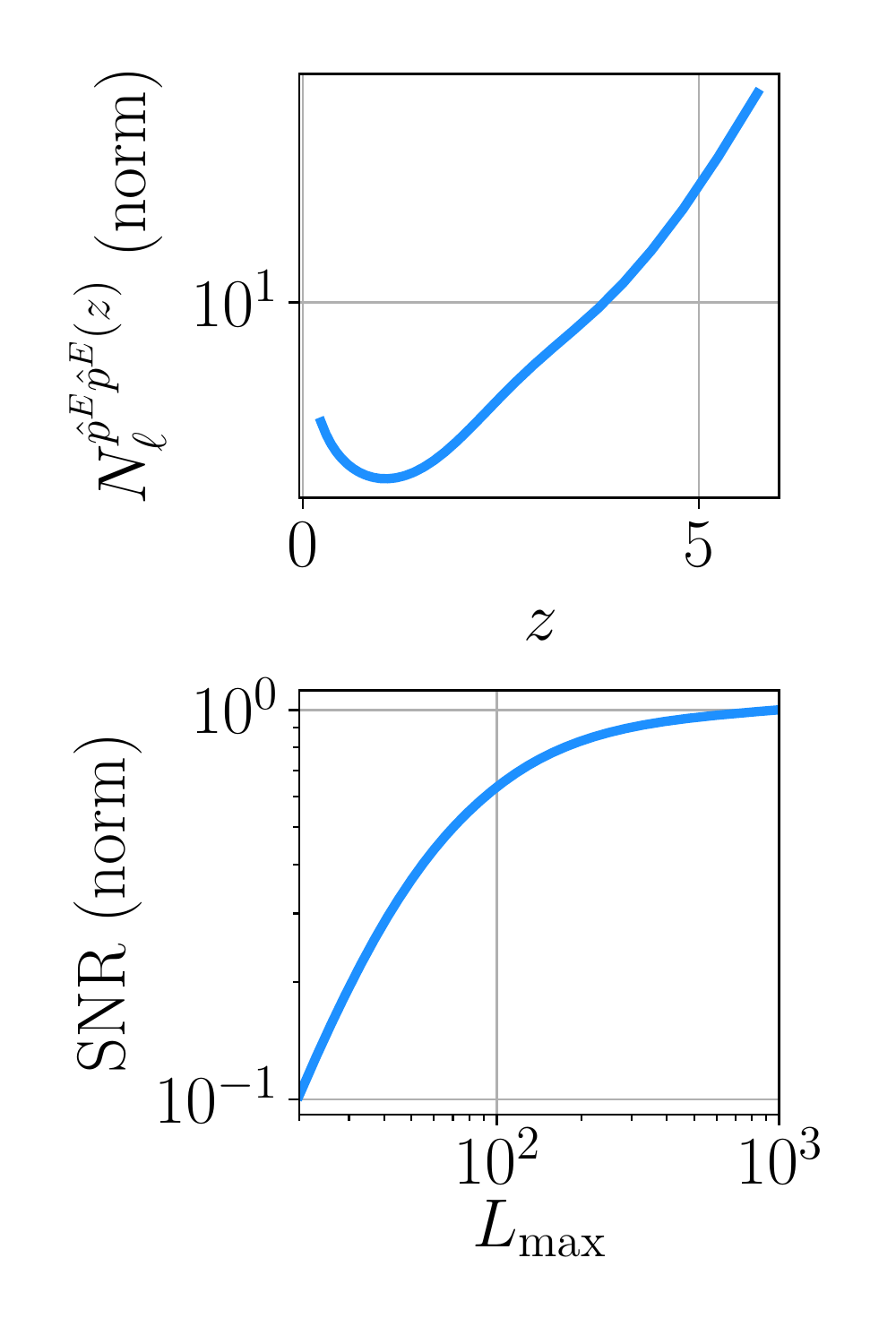}
    \vspace*{-0.8cm}
    \caption{(\textit{left}) The remote kinematic quadrupole power-spectra (solid blue line) shown together with the estimated reconstruction noise for combination of the VRO survey and two CMB surveys [CMB-S4 (dotted green) and CMB-HD (dashed red)] on small scales. We show the results for the fourth tomographic redshift bin where we take 32 bins spanning the range  $z\in[0.2,6.0]$ with the same radial size in comoving distance. (\textit{Top right}) The redshift dependence of reconstruction noise (normalised to its minimum value) for the same redshift bin. (\textit{Bottom right}) The cumulative SNR (normalised to its maximum value) for the same redshift bin as a function of the maximum multipole of the quadrupole reconstructed from the galaxy survey.}
    \label{fig:forecasts1}
\end{figure}

\begin{figure}[t!]
    \includegraphics[width=0.7\columnwidth]{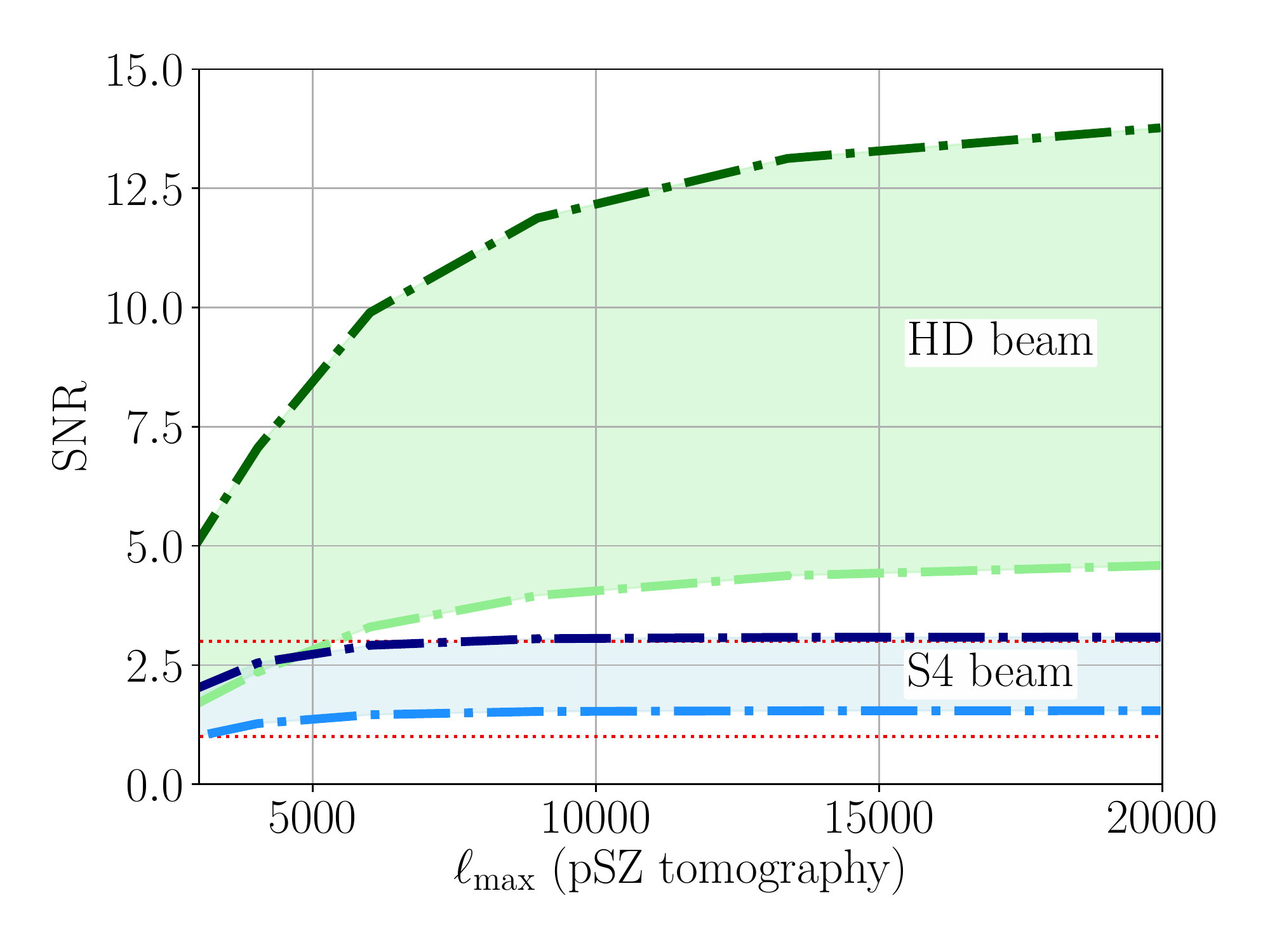}
    \vspace*{-0.8cm}
    \caption{\textit{The total detection SNR from the kpSZ tomography.} The dash-dotted green lines assume a CMB experiment whose specifications match the futuristic CMB-HD survey. The dash-dotted blue lines corresponds to the CMB-S4 survey. For both experiments, the upper darker-coloured dash-dotted lines correspond to assuming electron fluctuations trace dark matter on all scales, while lower lighter-coloured dash-dotted lines correspond to the standard `AGN' electron gas profile model. We take the baseline experimental configurations shown in Table~\ref{tab:beamnoise}. In both cases we take $f_{\rm sky}=0.5$ and show the SNR as a function of the maximum multipole included in the quadropole reconstruction. The top (bottom) red dotted line correspond to SNR value of 3 (1). Improving the beam of the CMB experiment allows better access to the non-Gaussian information on small-scales.}
    \label{fig:forecasts2}
\end{figure}

We  estimate the signal-to-noise ratio (SNR) of the reconstructed quadrupole assuming a cosmology consistent with the latest results from \textit{Planck}~\citep{Aghanim:2018eyx}. We perform the ILC cleaning taking into account the frequency-dependence of the kpSZ signal as described in Sec.~\ref{sec:ILC-cleaning}. We take two sets of baseline noise values for the CMB experiment which we describe in Table~\ref{tab:beamnoise} matching specifications anticipated from the CMB-S4 and the CMB-HD experiments. Note that the RMS noise values shown in Table~\ref{tab:beamnoise} are for temperature. We calculate the noise levels for polarization as $\Delta_E=\Delta_B=\sqrt{2}\Delta_T$ and model the CMB noise as in Eq.~\eqref{eq:CMB_noise}. We described our modelling of the galaxy field in Sec.~\ref{sec:direct_detection}. For the large-scale structure experiment we take the specifications of VRO~\citep{Abell:2009aa} and take 32 redshift bins in the range $z\in[0.2,6.0]$ with constant comoving bin width. Throughout, we model the electron distribution either by assuming the `AGN' gas profile from Ref.~\citep{Battaglia:2016xbi} using the ReCCO software~\citep{Cayuso:2021ljq}, or by assuming electrons trace dark matter on all scales. In the former case, physical processes such as AGN feedback, result in baryonic matter not tracing dark matter inside of halos, reducing the statistical power of pSZ tomography using galaxy tracers.

We compare the anticipated remote quadrupole signal from our fourth redshift bin to the reconstruction noise on the left plot in Fig.~\ref{fig:forecasts1}. The upper right plot in the same figure demonstrates the non-trivial dependence of the reconstruction noise on the decreasing density fluctuations and increasing visibility function $g(\eta)$ with increasing redshifts. The bottom right plot of this figure shows that the dominant contribution to the SNR comes from scales satisfying $\ell\sim100$.  In Fig.~\ref{fig:forecasts2}, we show the detection SNR from kpSZ tomography by combining the information from our 32 redshift bins as described above. We find a CMB-S4-~(CMB-HD) like survey can reach the SNR value of $\sim2~(\sim9)$ when combined with the VRO galaxy survey. Our analysis suggests that detection with CMB-S4 will likely be difficult while CMB-HD can potentially detect this signal to high SNR. Here, we omit calculating the bias on the velocity reconstruction from the primordial quadrupole assuming the ILC-cleaning will remove this significantly. Note also that the reconstructed kinetic quadrupole templates discussed throughout this work can potentially be used to reduce the contamination from kpSZ for the primordial gravitational wave searches by means of subtracting the kpSZ effect from CMB maps.

\section{Cosmology from the kpSZ effect }\label{sec:cosmology}

Measurement of the kpSZ effect may open new avenues for cosmological inference. Here, we give two possible applications and demonstrate the role of future kpSZ measurements in improving our understanding of the fundamental properties of the Universe.

\subsection{Detecting cosmic birefringence}

If the parity symmetry predicted by the standard LCDM is broken due to new physics, the plane of linear polarizations sourced due to pSZ effect can rotate due to the photon's coupling to the dark sector~\citep{Lue:1998mq,Carroll:1991zs,1992PhLB..289...67H}. We model this cosmic-birefringence effect on the $E$- and $B$- mode polarization due to kpSZ effect as 
\be
(Q'\pm iU')(\bn)=(Q\pm i U)^{\rm pkSZ}(\bn)e^{\pm 2i\alpha}
\ee
where the kpSZ superscript denotes the kinetic polarization signal sourced at the electron locations. We omit showing this superscript in what follows. From the above equation, the modulation of the polarization modes due to birefringence sourced between their emission and their arrival to our detectors is
\be
E_{\ell m}'&&=E_{\ell m}\cos2\alpha-B_{\ell m}\sin2\alpha \\ 
B_{\ell m}'&&=B_{\ell m}\cos2\alpha+E_{\ell m}\sin2\alpha\,.
\ee
The remote kinetic quadrupole fields measured from the cross of the CMB polarization and galaxies on small scales transforms similarly 
\be
{p^{E}_{\ell m}}' &&=p^{E}_{\ell m} \cos 2 \alpha-p^{B}_{\ell m} \sin 2 \alpha \\ 
{p^{B}_{\ell m}}'  &&=p^{B}_{\ell m}\cos2\alpha+p^{E}_{\ell m}\sin2\alpha\,.
\ee
Like before, we assume templates of $E$- and $B$-mode fluctuations induced by the kpSZ effect can be reconstructed from a large-scale structure survey. The set of observables from cross-correlations of the templates and the reconstructed remote quadrupole satisfy
\be
C_\ell^{{p}^E_i {p}^{E,\rm gal}_j}&\simeq& C_\ell^{{p}^E_i {p}_j^{E,\rm gal}}\cos(2(\alpha_i-\alpha_{\rm cal}))\,,\\
C_\ell^{{p}^{B}_i {p}^{B,\rm gal}_j}&\simeq& C_\ell^{{p}^{B}_i{p}^{B,\rm gal}_j}\cos(2(\alpha_i-\alpha_{\rm cal}))\,,\\
C_\ell^{{p}^{B}_i {p}^{E,\rm gal}_j}&\simeq& C_\ell^{{p}^{E}_i {p}_j^{E,\rm gal}}\sin(2(\alpha_i-\alpha_{\rm cal}))\,,\\
C_\ell^{{p}^E_i {p}^{B,\rm gal}_j}&\simeq&-C_\ell^{{p}^B_i {p}_j^{B,\rm gal}}\sin(2(\alpha_i-\alpha_{\rm cal}))\,,
\ee
where we parametrised the unknown CMB detector calibration with the parameter $\alpha_{\rm cal}$. Here, we only use the cross-correlation signal and define an ensemble information matrix as 
\begin{equation}
 \boldsymbol{\mathcal{F}}_{ab}=\sum_{\ell\ell'; ijkl} ({\partial C_\ell^{{p}^B_i\hat{p}^{E,\rm gal}_j}}/{\partial \boldsymbol{p}_a})\textbf{cov}^{-1}\left(\tilde{C}^{{p}^{B}_i\hat{p}^{E,\rm gal}_j}_{\ell},\tilde{C}^{{p}^B_k\hat{p}^{E,\rm gal}_l}_{\ell'}\right) ({\partial C_{\ell'}^{{p}_k\hat{p}^{E,\rm gal}_l}}/{\partial \boldsymbol{p}_b})  \, ,
\end{equation}
where $\boldsymbol{p}=\{\alpha_{\rm cal},\alpha\}$ are the set of parameters we consider, where we define $\alpha$ as the net rotation from a comoving range of $\alpha\simeq220$Mpc, assuming constant differential birefringence at all redshifts, i.e. assuming constant $\dd \alpha/\dd\chi$. 

\begin{figure}[t!]
    \includegraphics[width=0.7\columnwidth]{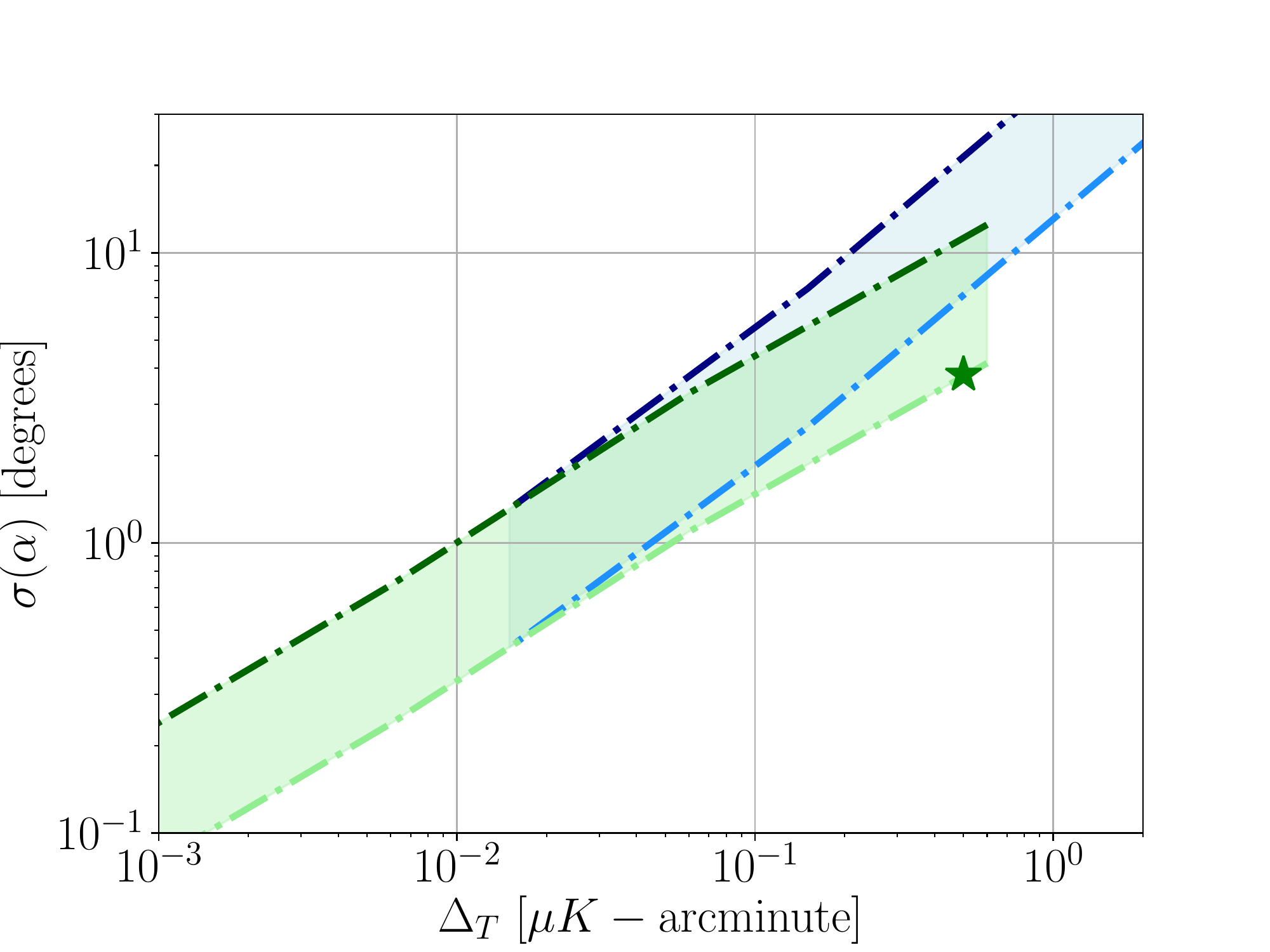}
    \vspace*{-0.4cm}
    \caption{\textit{Constraints on the birefringence angle per redshift bin of width $\Delta\chi~\sim220{\rm Mpc}$.}  We assume constant birefringence within each redshift bin of equal width $\Delta\chi$ in comoving distance, i.e. $\dot{\alpha}={\rm constant}$ where $\dot{\alpha}$ is the induced differential birefringence at comoving distance $\chi$, and we define $\alpha\equiv\int\dd\chi\dot{\alpha}$ with the integral in comoving distance spanning the range of each redshift bin. The blue-shaded region corresponds to constraints from the baseline CMB-S4 experiment we decribe in Table~I, but with varying RMS noise value $\Delta_T$. The {upper dark blue} line corresponds to modelling electrons with the `AGN' model, while the { lower light blue} line assumes electrons follow dark matter on all scales. The green-shaded region corresponds to the constraints from a CMB-HD like experiment with upper and lower lines defined similarly. We find that a CMB-S4 like experiment can potentially constrain birefringence angles as a function of redshift at $\sim10$ degree precision, while a CMB-HD like experiment can potentially reach degree-level constraints. The green marker corresponds to constraints using the current proposed experimental specifications of the CMB-HD experiment.}
    \vspace*{-0.2cm}
    \label{fig:birefringence}
\end{figure}

We demonstrate the prospect of measuring the birefringence angle for two CMB experiments in Fig.~\ref{fig:birefringence}, assuming VRO experimental specifications for the galaxy survey. We find that the kpSZ tomography can potentially detect rotation angle per redshift bin of width $\Delta \chi\sim220$Mpc to be $\sim1$ degree with future experiments. While these constraints may prove to be a marginal improvement compared to current and upcoming analyses e.g.~Refs.~\citep{Gruppuso:2011ci,Minami:2019ruj,Minami:2020fin,Diego-Palazuelos:2022dsq,Minami:2020odp, CMB-S4:2016ple}, they may provide a complementary probe of birefringence by breaking the calibration angle degeneracy through tomography, and by being subject to different systematics than constraints based on the recombination and reionization contributions to the CMB. Cross-correlation with the reionization $E$-modes, as well as the pSZ effect from the primordial quadrupole may potentially improve these constraints. Finally, note that the pSZ tomography from the primordial quadrupole signal can also probe cosmic-birefringence due to a similar degeneracy breaking. We expand on the observations made here in Ref.~\citep{Lee:2022udm}. Note that we have not considered the so-called `optical-depth degeneracy' due to mis-modelling of the electron profiles in the absence of perfect knowledge of the electron distribution~(see e.g. Ref.~\citep{Smith:2018bpn} for an in-depth discussion). This leads to a multiplicative bias on the reconstructed velocities. Here, we conjecture that a combination of the reconstructed remote quadruple auto-correlation along with cross-correlations with templates as discussed above, as well as the potential measurements of electron profiles (e.g. from fast-radio burst measurements~\citep{Madhavacheril:2019buy}), can mitigate the optical depth degeneracy.

\subsection{Primordial non-Gaussianity}

We now turn our attention to calculating the contribution to the remote quadrupole power-spectrum in the presence of a significant trispectrum, parameterized by an amplitude $\tau_{\rm NL}$, that can be sourced by models that deviate from the standard single-field slow roll inflation. Since the single-field slow roll inflation is just the simplest model, it is natural to wonder whether there may be other fields coupled to the inflaton. ``Cosmological collider physics''~\citep[e.g.][]{Baumann:2011nk,Assassi:2012zq,Chen:2012ge,Pi:2012gf,Noumi:2012vr,Arkani-Hamed:2015bza,Gong:2013sma,Lee:2016vti,Kehagias:2017cym,Kumar:2017ecc,An:2017hlx,An:2017rwo,Baumann:2017jvh,Kumar:2018jxz,Goon:2018fyu,Anninos:2019nib,Kumar:2019ebj,Hook:2019zxa} refers to the effort to seek the signatures of these new fields in the distribution of primordial density perturbations. One simple starting point in this inquiry is the coupling of the inflaton $\phi(x)$ to some other scalar $\psi(x)$ through an interaction $\propto\psi \phi^2$ (e.g., Ref.~\citep{Jeong:2012df}). Such a coupling would enhance the inflaton (and thus density, or curvature) four-point function (or trispectrum) without any enhancement in the three-point function (or bispectrum) \cite{2010PhRvD..82d3531T,Chen:2009zp,Suyama:2011qi}. A curvaton model could also enhance the trispectrum without much contribution to the bispectrum \citep{Tseliakhovich:2010kf,Smith:2010gx}. In many models, the trispectrum takes a ``local-model'' form, which amounts heuristically to a spatial modulation of the power-spectrum amplitude (equivalent to the scalar fossil field discussed in Ref.~\cite{Jeong:2012df}).  The {primordial} trispectrum amplitude $\tau_{\rm NL}$ is currently constrained to be $\tau_{\rm NL} \lesssim 2\!\times\!10^3$ at the 68\% confidence level by \textit{Planck}~\citep{Ade:2013ydc,Akrami:2019izv}. There is not much room to improve upon this significantly with CMB measurements  due to Silk damping of the temperature fluctuations. There are possible ways to seek $\tau_{\rm NL}$ in the trispectrum of the 21-cm brightness temperature \cite{Cooray:2008eb}, however, as well as the halo bias \cite{Ferraro:2014jba,Yamauchi:2015mja,Sekiguchi:2018kqe} and the 3-point correlations between two CMB temperature and one $\mu$ spectral distortion fluctuations \cite{Bartolo:2015fqz}; where the expected sensitivities are $\tau_{\rm NL}\sim 50 - 100$.

Locally sourced density perturbations produce fluctuations that scale as dictated by the Poisson equation, $\delta(\bk,z)\!\propto\!k^2\Phi(\bk)$, or with higher factors of $k$, hence are  much less significant at large scales compared to the primordial curvature fluctuations. Scale-dependence of {cosmological fluctuations} observed on the largest scales hence provide an opportunity to infer the initial conditions relatively unpolluted by the complicated local dynamics. Here, we are interested in a specific signal, the $\tau_{\rm NL}$-type non-Gaussianity, which boosts the signal from cross-correlations of fluctuations that are quadratic in density, such as the remote kinematic quadrupole field.  

The (connected) $N$-point correlation function of the gravitational potential is given by
\be
\langle\Phi(\bk_1)\!\ldots\!\Phi(\bk_N\!)\rangle_c\!=\!(2\pi)^3\delta_D(\bk_{1\dots N\!})F_\Phi(\bk_1,\ldots,\bk_N\!),\ \ \ \
\ee
where $\bk_{1\ldots N}=\sum_1^N\bk_{i}$ and $F_\Phi$ is a function that depends on the specific model of inflation. 
The collapsed 4-point function is equivalent to the large-scale correlation of small scale power, and therefore can be seen as directly related to the 2-point function of a quadratic {perturbation, as the one in defined in Eq.~\eqref{eq:remote_quad}, which can be generally expressed as}
\be\label{eq:intro_d2}
{\delta_2(\boldsymbol{x},z)=B(z)[\delta(\boldsymbol{x},z)]^2\,,}
\ee
{with 2-point correlations given by}
\be
{\langle\delta_2(\boldsymbol{x},z)\delta_2(\boldsymbol{x}',z')\rangle\!=\!B(z)B'(z')\!\langle[\delta(\boldsymbol{x},z)]^2[\delta(\boldsymbol{x}',z)]^2\rangle\,,}
\label{eq:d2d2}
\ee
and 
\be
\begin{split}
& \langle\delta_2(\bk_L,z)\delta_2'(\bk_L',z')\rangle =\!B(z)B'(z')\!\!\!\int_{\bk_1}\!\!\int_{\bk_3}\!\!\left\langle\delta(\bk_1,z)\delta(\bk_L\!-\!\bk_1,z)\delta'(\bk_3,z')\delta'(\bk_L'\!-\!\bk_3,z')\right\rangle\,,
\end{split}
\label{eq:d2d2_coll}
\ee
which can be seen as equal to the collapsed limit of the 4-point when $\bk_1\gg\bk_L$ and $\bk_3\gg\bk_L'$. Note also we defined $\int_{\bk}\equiv\int[\dd^3k/(2\pi)^3]$ {and $B(z)$ as a redshift dependent coefficient}. Eqs.~\eqref{eq:d2d2} and~\eqref{eq:d2d2_coll} show that two-point function of the remote quadrupole field $C_\ell^{pp}$ is a collapsed 4-point function.

In various multi-field inflation models \cite{2010PhRvD..82d3531T,Chen:2009zp,Suyama:2011qi}, as well as some curvaton models \citep{Tseliakhovich:2010kf,Smith:2010gx}, the observation signature of additional fields are well approximated by a spatial modulation of the power.  In this case, the four-point function for the gravitational potential becomes
\be
\lim\limits_{|\bk_{12}|\rightarrow0}F_{\Phi}(\bk_{1},\ldots,\bk_4)=16\left({5}/{6}\right)^2\tau_{\rm NL}P_{12}P_{1}P_{3},
\ee
where $P_i=P_\Phi(\bk_i)$ and $P_{ij}=P_\Phi(|\bk_i+\bk_j|)$.  This defines the non-Gaussian amplitude $\tau_{\rm NL}$ as commonly done in literature, and can be directly related to the amplitude of the local bispectrum, $f_{\rm NL}$, in the absence of additional degrees of freedom~(see e.g.,~\citep{Suyama:2007bg,Enqvist:2005pg,Byrnes:2006vq,Battefeld:2006sz,Vernizzi:2006ve,Seery:2006js,Kogo:2006kh, Choi:2007su, Enqvist:2008gk, Huang:2008bg, Chen:2009bc, Byrnes:2010em, 2010JCAP...04..009E, Smidt:2010ra}).

The contribution from primordial non-Gaussianity to the power spectrum of $\delta_2(\bk)$ field takes the form 
\begin{eqnarray}
\label{eq:power_ng_spec}
\left \langle\delta_2(\bk_L,z)\delta_2(\bk'_L,z)\right\rangle_{\rm NG}\!\!= 16 & &\,\left({5}/{6}\right)^2\!\tau_{\rm NL}\sigma^4 P_\Phi(k_L) \times (2\pi)^3\delta_D(\bk_L+\bk'_L)\,,\ \ 
\end{eqnarray}
where 
\be\sigma^2= B(z)\int\dd[\ln k/(4\pi)]\Delta^2_\delta({k},z)\ee 
is the variance of the density fluctuation. 

For the collapsed limit {and the inflationary models} described above,
\be
\begin{split}
\langle {\,_{\pm}}p(\bk_L) {\,_{\pm}}p(\bk_L')\rangle_{\rm NG}=f^2(x)(100/9)&\tau_{\rm NL}\sigma_{v_{\perp}}^4P_\Phi(\bk_L)\times(2\pi)^3\delta_D(\bk_L+\bk_L')\,.\ \ 
\end{split}
\ee
In order to calculate the non-Gaussian contribution to $C_\ell^{pp}$, we use the relation 
\be
(\delta_2)_{\ell m}=\int\dd^2\bn\, \delta_2(\chi\bn) Y_{\ell m}^*(\bn)\,,
\ee
to set
\be\label{eq:delta2lm}
(\delta_2)_{\ell m} =&& \int\dd^2\bn\int_{\bk}\delta_2(\bk,z)\Big[4\pi\sum\limits_{\ell'm'}i^{\ell'}j_{\ell'}(k\chi)Y_{\ell'm'}(\bn)Y_{\ell'm'}^*(\hat{\bk})\Big]Y_{\ell m }^*(\bn) \\ 
=&& \int_{\bk} \delta_2(\bk,z)\Big[\frac{16\pi^2}{2\ell+1}i^\ell j_\ell(k\chi)Y_{\ell m}^*(\bn)\Big]\,.
\ee
Using Eq.~\eqref{eq:delta2lm}, we write the non-Gaussian contribution to the collapsed 4-point function as
\be
\langle{\,_{\pm}}p_{\ell m}&&{\,_{\pm}}p^{*}_{\ell'm'}\rangle_{\rm NG}\\
&&=\int_{\bk}\int_{\bk'}\langle\delta_p(\bk,z)\delta_p(\bk',z)\rangle_{\rm NG}\Big[\frac{16\pi^2}{2L+1}i^L j_L(k\chi) Y_{LM}^*(\hat{\bk})\Big]\Big[\frac{16\pi^2}{2L'+1}i^{L'} j_{L'}(k'\chi) Y_{L'M'}^*(\hat{\bk}')\Big]\nonumber \\
&&=f^2(x)(100/9)\tau_{\rm NL}\,\sigma_{v_\perp}^4\!\!\int\frac{k^2 \dd k}{(2\pi)^3}[16\pi^2 j_L(k\chi)]^2\frac{4\pi}{(2L+1)^3}P_{\Phi}(k)\,,
\ee
where 
\be P_{\Phi}(k)=\frac{2\pi^2}{k^3}A_s\Big(\frac{k}{k_\star}\Big)^{1-n_s}\,.\ee 
Here, $A_s=2.2\times10^{-9}$ and $n_s=0.965$ for $k_\star=0.5/{\rm Mpc}$. We define the transverse velocity variance as
\be
\sigma^2_{v_\perp}= \int\dd[\ln k/(4\pi)]W(k\chi)^2\Delta^2_{v_\perp}({k},z)
\ee 
which is equivalent to the RMS value of the bulk transverse velocity, defined with a window function $W(k\chi)$ which sets the maximum Fourier wavenumber $k$ of the integral. In what follows we will take the RMS velocity to satisfy $\sigma_{\rm v_\perp}=2\!\times\!10^{-3}c\approx600$km$/$s. 

\begin{figure}[t!]
    \includegraphics[width=0.7\columnwidth]{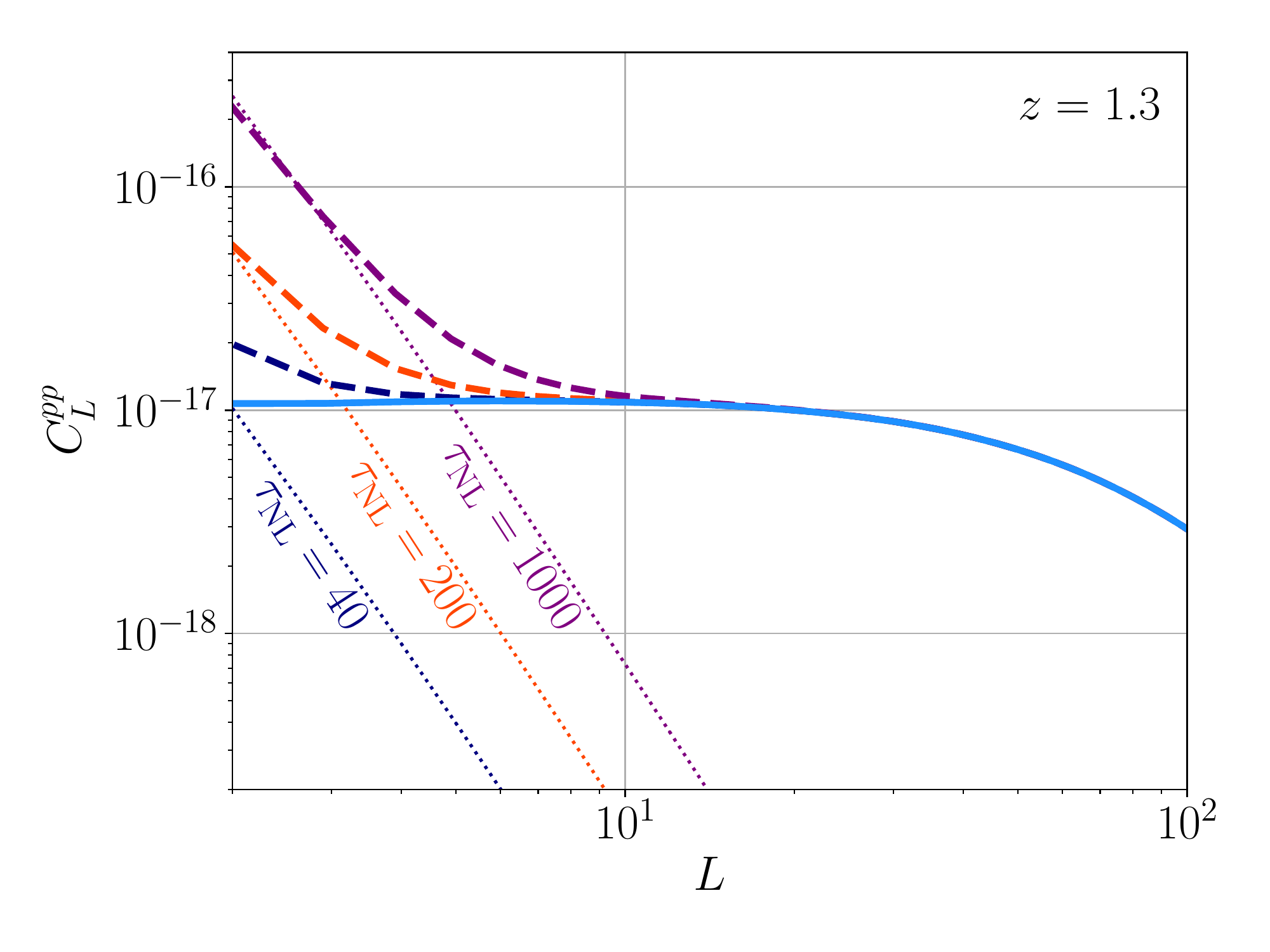}
    \vspace*{-0.9cm}
    \caption{The enhancement of the  kinematic remote quadrupole signal due to $\tau_{\rm NL}$ non-Gaussianity. The dotted lines show the non-Gaussianity signal for $\tau_{\rm NL}$ values of $\{40,200,1000\}$ from bottom to top. We see on large-scales the non-Gaussian contribution to the kinematic remote quadrupole power-spectra indeed dominates over the Gaussian contribution.}
    \label{fig:forecasts3}
\end{figure}

We show the non-Gaussian contribution to the kinetic remote quadrupole power-spectra in Fig.~\ref{fig:forecasts3}. We find a CMB-S4 (CMB-HD) like experiment can detect $\tau_{\rm NL}$ value of 400 (100) at $68\%$ confidence if electrons trace matter fluctuations at all scales. Assuming an electron gas model, we find the lowest value of $\tau_{\rm NL}$ that can be detected at $68\%$ confidence is $\sim1000$ ($\sim300$) with CMB-S4 (CMB-HD). While these values are less impressive compared to the statistical power of galaxy surveys~\citep{Ferraro:2014jba} or kSZ tomography~\citep{AnilKumar:2022flx}, for example, they are nevertheless an order of magnitude smaller than the current constraints from the CMB, and can potentially play a role in validating findings using different methods. 

\begin{figure}[t!]
    \includegraphics[width=0.7\columnwidth]{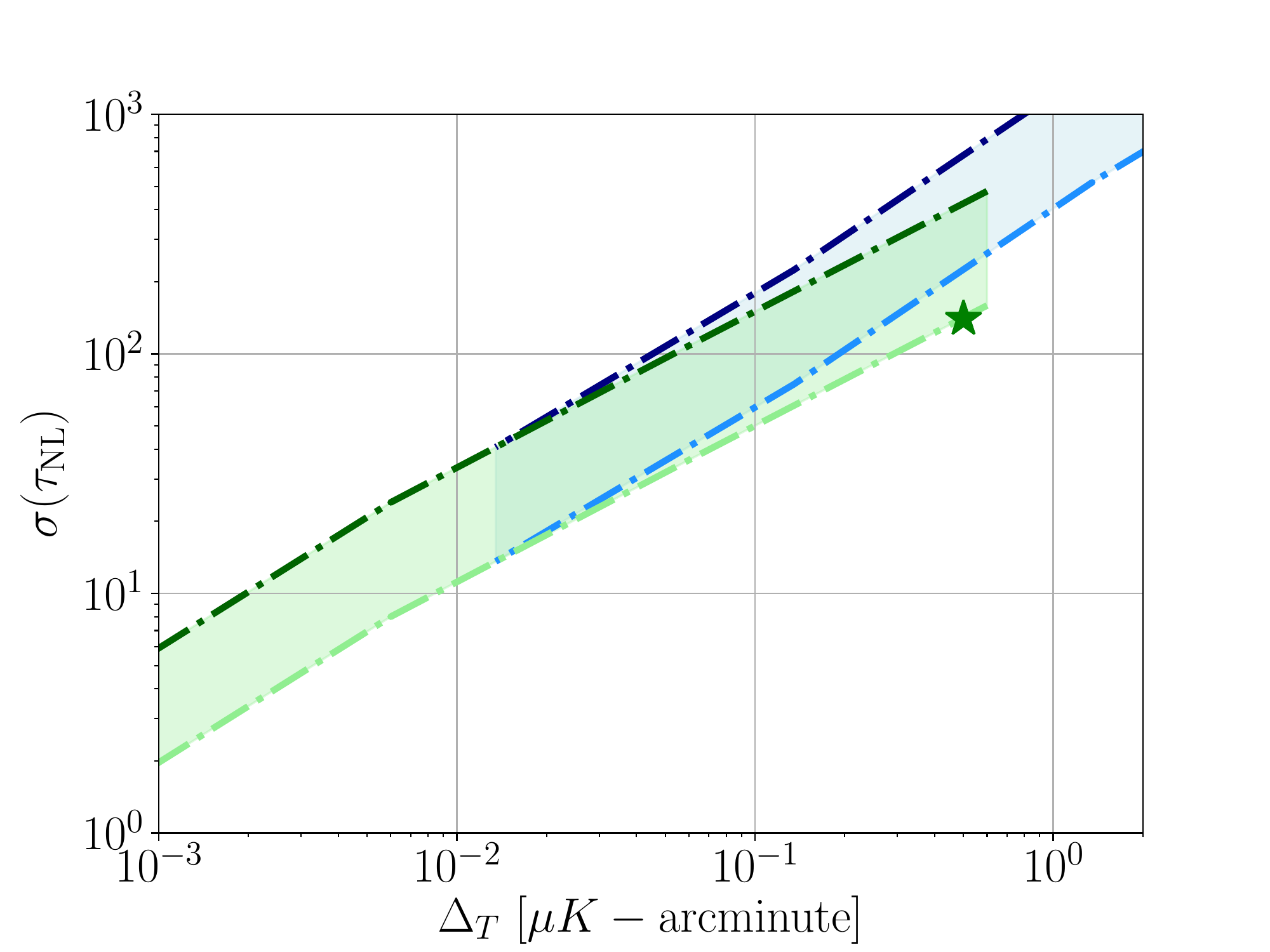}
    \vspace*{-0.4cm}
    \caption{Constraints on the amplitude of the collapsed 4-point non-Gaussianity, $\tau_{\rm NL}$ for a range of CMB experiment RMS values with other experimental specifications matching the baseline configuration of Table~I.  The upper {dark} blue line corresponds to modelling electron fluctuation using the AGN model, while the lower {light} blue line assumes electrons follow matter fluctuations on all scales. The green-shaded region corresponds to the constraints from a CMB-HD like experiment with upper and lower lines defined similarly. We find that a CMB-S4 like experiment can potentially constrain primordial non-Gaussianity of the $\tau_{\rm NL}$ type down to $\tau_{\rm NL}\sim400$ precision, while a CMB-HD like experiment can potentially improve these constraints by a factor of $\sim4$. The green marker corresponds to constraints using the current proposed experimental specifications of the CMB-HD experiment.}
    \label{fig:forecasts4}
\end{figure}

\section{Discussion}\label{sec:conclusions}

The next generation of CMB experiments and galaxy surveys are poised to drive a paradigm shift, in which a wide variety of secondary effects in the CMB are detected at high enough significance to become powerful cosmological probes in their own right. In this paper, we discussed the prospects for these experiments to detect the kinetic polarized Sunyaev Zel'dovich (kpSZ) effect using two different techniques. First, we demonstrated that on large (roughly degree) angular scales it is possible to detect the kpSZ effect by using multifrequency measurements of the CMB from CMB-S4, CMB-HD, or PIXIE in cross-correlation with a transverse velocity template constructed using a Rubin LSST-like galaxy survey. Next, we constructed a quadratic estimator for the kinetic component of the quadrupole field based on the anisotropic small-angular scale cross-correlation between CMB and a galaxy survey. We demonstrated how the reconstructed remote kinetic quadrupole field can be used to constrain cosmic birefringence and primordial non-Gaussianity. 

In future work, it will be important to compare the forecasts presented here to the analysis of simulations, where the full non-linear effects of gravity are accounted for. The potential biasing of the reconstruction due to mis-modelling of the electron density field, as well as the likely enhancement of noise due to this foreground when using different methods such as real-space matched filtering, also warrant further analysis. Despite the potential hurdles, the outlook for detecting kpSZ with near-term experiments is promising, and it may provide novel constraints on various cosmological scenarios. The kpSZ effect is just one of a number of CMB secondaries, and the detection of this signal will undoubtedly be an important milestone towards the development of new cosmological observables and new methods to constrain fundamental physics with future experiments.

\section{Acknowledgements}

We thank Jose Bernal Luis, Jens Chuluba, Simone Ferraro, Andrew~H.~Jaffe, Nanoom Lee and Joel Meyers for useful conversations. SCH thanks Jose Bernal Luis for useful discussions on the conceptualization of the non-Gaussian $\tau_{\rm NL}$ signal and Andrew~H.~Jaffe for their contributions at the early stages of this project. This work was started in part at the Aspen Center for Physics, which is supported by National Science Foundation grant PHY-1607611. SCH is supported by the Horizon Fellowship from Johns Hopkins University. SCH also acknowledges the support of a grant from the Simons Foundation at the Aspen Center for Physics. MCJ is supported by the National Science and Engineering Research Council through a Discovery grant. This research was supported in part by Perimeter Institute for Theoretical Physics. Research at Perimeter Institute is supported by the Government of Canada through the Department of Innovation, Science and Economic Development Canada and by the Province of Ontario through the Ministry of Research, Innovation and Science.

\bibliography{main}

\end{document}